\documentclass[epj]{svjour}
\usepackage{amsmath,amssymb,graphicx,bm}
\usepackage{color}
\usepackage{amsmath,amssymb,bm,natbib}
\usepackage{epsfig}
\usepackage{graphicx}
\usepackage{slashed}

\newcommand{\beq}{\begin{equation}}
\newcommand{\eeq}{\end{equation}}
\newcommand{\bea}{\begin{eqnarray}}
\newcommand{\eea}{\end{eqnarray}}
\newcommand{\ba}{\begin{align}}
\newcommand{\ea}{\end{align}}
\newcommand{\bfig}{\begin{figure}}
\newcommand{\efig}{\end{figure}}

\newcommand{\D}{\displaystyle}

\newcommand{\gev}{\, \text{GeV}}
\newcommand{\mev}{\, \text{MeV}}

\newcommand{\tin}{t_{\rm in}}
\newcommand{\la}{\langle}
\newcommand{\ra}{\rangle}

\newcommand{\omnes}{{\cal{O}}}

\begin{document}

\vspace{1cm}

\title{Model independent bounds on the modulus of the pion form factor on the unitarity cut  below the $\omega\pi$ threshold}
\author{B. Ananthanarayan$^{a}$\and I. Caprini$^{b}$ \and Diganta Das$^{c}$ 
\and I. Sentitemsu Imsong$^{a}$ }
\institute{$^a$ Centre for High Energy Physics,
Indian Institute of 
Science, Bangalore 560 012, India \\
$^b$ Horia Hulubei National Institute for Physics and Nuclear Engineering,
P.O.B. MG-6, 077125 Magurele, Romania\\
$^c$ Institute of Mathematical Sciences,
 Taramani, Chennai 600113, India}


\abstract{We calculate upper and lower bounds on the modulus of the pion electromagnetic form factor on the 
unitarity cut below the  $\omega\pi$ inelastic threshold,  using as input the phase in the elastic region  
known  via the Fermi-Watson theorem from the $\pi\pi$ $P$-wave phase shift,  and a suitably weighted integral 
of the modulus squared above the inelastic threshold.  The normalization at $t=0$,  the pion charge radius and 
experimental values at spacelike momenta are used as additional input information. The bounds are model independent, 
in the sense that they do not rely on  specific parametrizations and do not require assumptions on the phase of the form factor above the inelastic 
threshold.  The results provide nontrivial consistency
checks on the recent experimental data on the 
modulus available below the $\omega\pi$ threshold from $e^+ e^-$ annihilation and $\tau$-decay experiments. 
In particular, at low energies the calculated bounds offer a more precise description of the modulus than 
the experimental data. }


\titlerunning{Bounds on the pion form factor}

\authorrunning{B. Ananthanarayan et al.} 
\maketitle
\section{Introduction}
\label{sec:intro}
The pion  electromagnetic form factor $F(t)$ is a fundamental observable 
that has been used as a probe of QCD in several regimes, including  
Chiral Perturbation Theory (ChPT) and lattice simulations at low energies and perturbative QCD at
asymptotic spacelike momenta. 
On the other hand, even before the rise of QCD as the modern theory of strong
interactions, form factors were analyzed based on the
general principles of analyticity and unitarity.  

The experimental information on the pion form factor improved significantly in recent years. In particular, 
the modulus $|F(t)|$ on the unitarity cut $t> t_+$, where
$t_+ = 4 M_\pi^2$, was  measured by several  $e^+e^- \to \pi^+\pi^-$ experiments \cite{BABAR, KLOE1, KLOE2, CMD2:1, CMD2:2} 
and, invoking isospin symmetry, also  from hadronic $\tau$-decays \cite{Fujikawa:2008ma}.  
In the elastic region $t_+ \leq t \leq t_{\rm in}$ where $t_{\rm in} = (M_\omega + M_{\pi^0})^2$ 
is the first significant inelastic threshold, the phase is known via the Fermi-Watson theorem 
from the phase-shift of the $P$-wave $\pi\pi$ elastic scattering amplitude, determined recently 
with great accuracy from Roy equations and scattering data  \cite{ACGL,CGL,KPY,GarciaMartin:2011cn,Caprini:2011ky}.  
While an obvious
correlation between the phase and modulus is implied by explicit formulae such as Breit-Wigner parametrizations,
independent sources of correlation are required to test these measurements in a model independent way.

In this work we present  upper and lower bounds on the
modulus of the form factor in the elastic region, resulting from phase information on the 
unitarity cut in the elastic region. 
Additional inputs coming from the normalization at $t=0$, the knowledge of the pion charge radius 
\cite{Colangelo:2004, Masjuan} and spacelike measurements \cite{Horn, Huber} can
offer further constraints. The appropriate tools for this investigation are analyticity
and unitarity, which correlate the various pieces of the input. Recently developed mathematical 
techniques \cite{IC, Abbas:2010EPJA, AC}, which
were used to constrain the pion form factor near the origin \cite{Ananthanarayan:2011xt} 
and on the spacelike axis \cite{ Ananthanarayan:2012tn}, can be readily extended to 
the elastic region of the unitarity cut.  
 
 The method uses as first input  the knowledge of the phase 
\beq\label{eq:watson}
{\rm Arg} [F(t+i\epsilon)]=\delta_1^1(t), \quad\quad  t_+ \leq t \leq \tin,
\eeq
where $\delta_1^1(t)$ is the phase shift of the $P$-wave of $\pi\pi$ elastic scattering. 

The second piece of input comes from the knowledge of $|F(t)|$ above $\tin$.  The modulus has been measured up to  $\sqrt{t}=3\, \gev$  by BaBar Collaboration \cite{BABAR}. Combined with the asymptotic decrease
$|F(t)|\sim 1/t$  at large $t>0$, which can be inferred from the perturbative QCD prediction at large space like momenta and 
 mathematical theorems on the asymptotic behaviour of analytic functions \cite{Cornille:1975}, this information can be used to calculate an integral defined as
 \beq\label{eq:L2}
 \D\frac{1}{\pi} \int_{\tin}^{\infty} dt \rho(t) |F(t)|^2 = I.
 \eeq
Here $\rho(t)$ is a positive-definite weight, for which the integral converges and an accurate evaluation of $I$ is possible.
The optimal procedure is to vary $\rho(t)$ over 
a sufficiently large admissible class and take the best result. A suitable  class is given by the simple expression
\beq \label{eq:rhogeneric0}
\rho(t) = \frac{t^b}{(t+Q^2)^c}, \quad \quad 
\eeq
where  $Q^2>0$ and $b, c$ taken in the range $b\leq c \leq b+2$. 
The dimension of $I$ can be inferred from Eqns.(\ref{eq:L2}) and (\ref{eq:rhogeneric0}).
As we shall discuss, in the choice of 
$Q^2, b, c$ a compromise should be achieved between the objective of deriving sufficiently strong bounds and the 
need of a precise evaluation of the integral (\ref{eq:L2}) with the present knowledge of $|F(t)|$.

Additional information inside the analyticity domain can also be implemented exactly.
In practice we shall use the normalization and the first derivative at $t=0$:
\beq\label{eq:taylor}
F(0)=1, \quad \quad \quad  F'(0)=\frac{1}{6} \la r_\pi^2 \ra,
\eeq
with the charge radius $\la r_\pi^2 \ra$ varied within reasonable limits  \cite{Colangelo:2004, Masjuan}. 
We shall further implement  the value
of the form factor at a spacelike momentum $t_s<0$,
\beq\label{eq:spacelike}	
F(t_s)= F_s\pm \epsilon_s,
\eeq
known from the most precise experiments \cite{Horn, Huber}.

The relations (\ref{eq:watson}),  (\ref{eq:L2}), (\ref{eq:taylor}) and (\ref{eq:spacelike}) 
define a class of functions $F(t)$ real analytic  in the $t$-plane cut for $t>t_+$. In Refs. 
\cite{Ananthanarayan:2011xt, Ananthanarayan:2012tn}, bounds on the higher 
Taylor coefficients at $t=0$ and  on the values of $F(t)$ on the spacelike axis, 
for  functions $F(t)$ belonging to this class  were derived.
We extend now the formalism and derive rigorous upper and lower bounds on the 
modulus $|F(t)|$ in the region $t_+<t<\tin$, by solving the corresponding 
extremal problem on the same functional class. These bounds 
will be compared with 
recent experimental measurements of the modulus from  $e^+e^-\to \pi^+\pi^-$ 
\cite{BABAR, KLOE1, KLOE2, CMD2:1, CMD2:2}  and  
$\tau$-decays \cite{Fujikawa:2008ma}
below the $\omega\pi$ threshold.

We emphasize that the goal of present work is to check (and confirm) 
the consistency of various data sets on the modulus with the bounds 
derived from the input described above. 
Therefore, we have not used as input in the formalism 
the values of the modulus below  the inelastic threshold provided by 
the recent experiments. 
Adding to the input one or more of these measured values
would lead to a considerable strengthening
of the bounds. However, as the different sets of data are not 
perfectly consistent amongst themselves, the choice 
of the input from one set would introduce a certain bias, 
which we prefer to avoid in this first analysis. 
It may also be noted that one could use the techniques
applied here to improve, for instance, the extracted value of the pion radius, 
using the precise values of the recently measured modulus. 
These problems are of interest and will be considered in a future work.

The scheme of the paper is as follows:  in Sec. \ref{sec:method},
we give a brief review of the mathematical formalism. In Sec. \ref{sec:inputs}, we describe
the  information used as input, discussing also the inclusion of the main isospin breaking effect, 
produced by the $\rho-\omega$ interference, which affects the form factor extracted from $e^+e^-$ annihilation. 
In Sec. \ref{sec:results},  we present the results of
our computations, including a detailed analysis of the  uncertainties of the input, and make a  
comparison with the most recent experimental measurements of the modulus from  $e^+e^-\to \pi^+\pi^-$ 
\cite{BABAR, KLOE1, KLOE2, CMD2:1, CMD2:2}  and  $\tau$-decays \cite{Fujikawa:2008ma}.  Finally 
in Sec. \ref{sec:conclusions}, we provide a discussion and a summary of our results. 
\section{Method}
\label{sec:method}

We exploit (\ref{eq:watson}) by introducing the Omn\`es function defined by
\beq	\label{eq:omnes}
 \omnes(t) = \exp \left(\D\frac {t} {\pi} \int^{\infty}_{t_+} dt' 
\D\frac{\delta (t^\prime)} {t^\prime (t^\prime -t)}\right),
\eeq
where $\delta(t)=\delta_1^1(t)$   for
$t\le \tin$, and is an arbitrary function, sufficiently  smooth ({\em i.e.,}
Lipschitz continuous) for $t>\tin$.

The crucial remark is that the function $h(t)$, defined by
\beq\label{eq:h}
F(t)= \omnes(t) h(t),
\eeq
is real for $t<\tin$, {\em i.e.} it is analytic in the $t$-plane cut only along $t>\tin$. 
Moreover, from (\ref{eq:L2}) it follows that $h(t)$  satisfies the condition 
\beq\label{eq:hL2}
\D\frac{1}{\pi} \int_{\tin}^{\infty} dt\, 
\rho(t) |\omnes(t)|^2 |h(t)|^2 = I.
\eeq
In order to exploit this property, we cast the problem into a canonical form by performing the conformal transformation
\beq\label{eq:ztin}
\tilde z(t) = \frac{\sqrt{\tin} - \sqrt {\tin -t}} {\sqrt{\tin} + \sqrt {\tin -t}}\,,
\eeq
which maps the complex $t$-plane cut for $t>t_{\rm in}$ onto the unit disk $|z|<1$
in the $z$-plane defined by $z \equiv \tilde z(t)$.
The upper (lower) lip of the branch-cut $[\tin, \infty]$ is mapped onto the
upper (lower) half of the unit circle in the complex $z$-plane, the real line $[-\infty,0]$ to $[-1,0]$
and $[0,\tin]$ to $[0,1]$. 

The next step is to introduce two outer functions, {\it i.e.} functions analytic and without zeros in 
the unit disk $|z|<1$, defined in terms of their modulus on the boundary, related to  $\sqrt{\rho(t)\, |{\rm d}t/ {\rm d} \tilde z(t)|}$ 
and  $|\omnes(t)|$,  respectively \cite{IC, Abbas:2010EPJA}.
In particular, for weight functions with the expression (\ref{eq:rhogeneric0}), the first outer function $w(z)$  
can be written in an analytic closed form in the $z$-variable as \cite{Abbas:2010EPJA}
\beq\label{eq:outerfinal0}
w(z)= (2\sqrt{t_{\rm in}})^{1+b-c}\frac{(1-z)^{1/2}} {(1+z)^{3/2-c+b}}\frac{(1+\tilde z(-Q^2))^c}{(1-z \tilde z(-Q^2))^c},
\eeq 
where $\tilde z(-Q^2)$ is computed using Eq.  (\ref{eq:ztin}).

For the second outer function, denoted as $\omega(z)$,  we shall use an integral representation in terms 
of its modulus on the cut $t>\tin$, which can be written as \cite{IC, Abbas:2010EPJA}
\beq\label{eq:omega}
 \omega(z) =  \exp \left(\D\frac {\sqrt {\tin - \tilde t(z)}} {\pi} \int^{\infty}_{\tin}  \D\frac {\ln |\omnes(t^\prime)|\, {\rm d}t^\prime}
 {\sqrt {t^\prime - \tin} (t^\prime -\tilde t(z))} \right),
\eeq 
where $\tilde t(z)$ is the inverse of $z = \tilde z(t)$, for $\tilde z(t)$  defined in
(\ref{eq:ztin}). We mention that the product $w(z)\, \omega(z)$  defines in fact a single outer function with 
modulus accounting for all the factors in (\ref{eq:hL2}). The separation into two functions was suitable in our case since one of them, $w(z)$, could be written in the simple analytic form  (\ref{eq:outerfinal0}).

Further, if we define a function $g(z)$ by
\beq\label{eq:gF}
 g(z) = w(z)\, \omega(z) \,F(\tilde t(z)) \,[\omnes(\tilde t(z)) ]^{-1}, 
\eeq 
 it is easy to see that the integral
(\ref{eq:hL2}) can be written in terms of $|g(z)|^2$ for $z$ on the boundary ($z={\rm e}^{i\theta}$) as
\beq\label{eq:gI1}
\frac{1}{2 \pi} \int^{2\pi}_{0} {\rm d} \theta |g({\rm e}^{i\theta})|^2 = I.
\eeq
The  $L^2$-norm condition (\ref{eq:gI1}) leads to what is known as the Meiman interpolation problem, 
which consists in finding the most general rigorous correlations between the values of the function and 
its derivatives inside the unit disk $|z|<1$. These conditions can be written either as a positivity 
of a determinant and its minors, or as an inequality for an explicit convex quadratic expression of the input quantities
(for a proof and older 
references see \cite{Abbas:2010EPJA}). For instance, if we denote $g_0=g(0)$ and $g_1=g'(0)$ the value of the function $g(z)$ and the first derivative at $z=0$, and by $g(z_a)$ and $g(z_b)$ the real values at two real points  $|z_a|<1$ and $|z_b|<1$, then (\ref{eq:gI1}) implies the determinantal inequality 
\beq\label{eq:det2}
\left|
\begin{array}{c c c}\vspace{0.2cm}
I-g_0^2-g_1^2 & \bar g(z_a)  & \bar g(z_b)\\\vspace{0.2cm}	
	\bar g(z_a)  & \D \frac{z^{4}_{a}}{1-z^{2}_a} & \D
\frac{(z_a z_b)^2}{1-z_a z_b} \\
	\bar g(z_b)  &\D \frac{(z_a z_b)^{2}}{1-z_a z_b} & 
\D \frac{(z_b)^{4}}{1-z_b^2}\\
	\end{array}\right| \ge 0,
\eeq
where
\beq
 \bar g(z_a)=g(z_a) - g_0-g_1 z_a, \quad \bar g(z_b)= g(z_b) - g_0-g_1 z_b.\eeq
Furthermore, all the minors of the determinant  above should be nonnegative.

In the present case we note that  (\ref{eq:ztin}) implies that the origin $t=0$ of the $t$-plane is
mapped onto the origin $z=0$ of the $z$-plane. Then the values $g_0$ and $g_1$ are related in a 
straightforward way to  $F(0)=1$ and the charge radius $\la r_\pi^2 \ra$.
 Also, we shall take $z_a=\tilde z(t_a)<0$, with $t_a=t_s$  a spacelike point  (\ref{eq:spacelike}) 
included as input, and  $0<z_b=\tilde z(t_b)<1$, where $t_b$ is an arbitrary point in the range $(t_+, \,\tin)$.  
Then using as input $F(t_a)= F(t_s)$ from (\ref{eq:spacelike}), {\em i.e.} a known value for
\beq\label{eq:xin}
g(z_a)=w(z_a)\, \omega(z_a) \,F(t_a) \,[\omnes(t_a) ]^{-1}, \quad z_a=\tilde z(t_a),
\eeq
 the inequality (\ref{eq:det2}) leads to upper and lower bounds for the real values $g(z_b)$. 
From these we obtain upper and lower bounds on the modulus $|F(t_b)|$, written using (\ref{eq:gF}) as
 \beq\label{Fg}
|F(t_b)|= |\omnes(t_b)|\, \frac{g(\tilde z(t_b))}{ w(\tilde z(t_b))\, \omega(\tilde z(t_b))}.
\eeq
We recall that $|\omnes(t)|$ is obtained from (\ref{eq:omnes}) via the Principal Value ($PV$) Cauchy integral
\beq	\label{eq:modomnes}
 |\omnes(t)| = \exp \left(\frac {t} {\pi} PV \int^{\infty}_{4 M_\pi^2} dt' 
\D\frac{\delta (t^\prime)} {t^\prime (t^\prime -t)}\right).
\eeq
It is important to emphasize that, as proven in  \cite{IC, Abbas:2010EPJA}, the bounds are independent of the  
phase $\delta(t)$ for $t>\tin$ entering the Omn\`es function (\ref{eq:omnes}). 
Also, it is easy to show \cite{IC,Abbas:2010EPJA} that, at fixed
$\rho(t)$, the  bounds  depend in a monotonous way 
on the value of the quantity $I$, becoming stronger when this 
value is decreased.

\section{Input}\label{sec:inputs}
 Recall that the first inelastic threshold $\tin$ is due to the opening of the $\omega\pi$ channel, $\sqrt{\tin}=0.917\,\gev$. 
Below $\tin$ we use as phenomenological input the phase shift  $\delta_1^1(t)$ determined recently with high precision from  Roy equations satisfied
by the $\pi\pi$ elastic amplitude  \cite{ACGL,CGL, KPY,GarciaMartin:2011cn,Caprini:2011ky}. 
In particular,  the $P$-wave phase-shift is parametrized in \cite{GarciaMartin:2011cn} as
\beq\label{eq:delta11}
{\rm cot}\delta_1^1(t)=\frac{\sqrt{t}}{2k^3}(M_\rho^2 - t) \!\!\left(\frac{2 M_\pi^3}{M_\rho^2 \sqrt{t}} + B_0 + 
      B_1 \frac{\sqrt{t} -\sqrt{t_{0} - t}}{\sqrt{t} + \sqrt{t_{0} - t}}\!\!\right)\!, 
\eeq
where $k=\sqrt{t/4 -M_\pi^2}$ and  $\sqrt{t_0}=1.05\, \gev$,  $B_0=1.043\pm 0.011$, $B_1=0.19 \pm 0.05$ and $M_\rho= 773.6 \pm 0.9\, \mev$ 
(we have taken the parameters from the so-called CDF solution \cite{GarciaMartin:2011cn}).

The phase  $\delta_1^1(t)$ obtained from (\ref{eq:delta11}) with the central values of the parameters 
is  very close to the solution of the Roy equations presented recently in \cite{Caprini:2011ky}: 
the differences between the two phases are less than a small fraction of a degree at low energies and 
reach one degree only at the upper end $\tin$ of the elastic range. In fact, these differences are 
comparable with the  uncertainty resulting from the errors of the parameters quoted below (\ref{eq:delta11}). 
In our analysis we have used for the central value of the phase the expression (\ref{eq:delta11}) and estimated 
the  uncertainty either from the errors of the parameters quoted above or as the difference between the 
phase (\ref{eq:delta11}) and the phase obtained in \cite{Caprini:2011ky}.

The phase shift given above was obtained  assuming isospin symmetry, with $M_{\pi}=M_{\pi^+}$.
As discussed in \cite{Leutwyler:2002hm}, an important isospin breaking effect, which manifests itself in $e^+e^-$ 
annihilation, is produced by $\rho-\omega$ interference. We draw attention also on the extensive analyses of the data from $e^+e^-$ annihilation and $\tau$-decay,  in particular of the isospin breaking effects,  performed
recently in \cite{Jegerlehner:2011ti,Benayoun:2011mm}, while the comparison of experiment and theory in $e^+e^-\to$ hadrons was discussed in detail in \cite{Jegerlehner:2009ry,Davier}, in connection with the hadronic contribution
to the anomalous magnetic moment of the muon $(g-2)$.

In our work we shall model the isospin breaking produced by  $\rho-\omega$ interference effect by adding to the phase 
$\delta(t)$ entering the Omn\`es function  (\ref{eq:omnes}) the correction
$\Delta \delta(t)={\rm Arg} \left[F_{\rho-\omega}(t) \right]$, where \cite{Leutwyler:2002hm, Hanhart:2012wi}
\begin{equation} \label{eq:iso}
F_{\rho-\omega}(t)=\Big(1+\epsilon\,\frac{t}{t_\omega-t} \Big), ~~ t_\omega=(M_\omega-i/2 \Gamma_\omega)^2, 
\end{equation}
with  $M_\omega=0.7826\,\gev$, $\Gamma_\omega=0.0085\,\gev$, and the small parameter $\epsilon=1.9\times 10^{-3}$. 

Alternatively, the effect can be accounted by multiplying the bounds derived in the isospin limit with the modulus 
of the correction factor $F_{\rho-\omega}(t)$ given in (\ref{eq:iso}). The two procedures give practically the same results.
 
Above $\tin=(0.917\, \gev)^2$  we use in (\ref{eq:omnes}) a continuous function $\delta(t)$, 
which approaches asymptotically $\pi$. As shown in \cite{Abbas:2010EPJA}, if 
this function is Lipschitz continuous, the dependence on $\delta(t)$ of the functions $\omnes(t)$ and $\omega(z)$,  
defined in (\ref{eq:omnes}) and (\ref{eq:omega}), respectively, exactly compensate  each other, leading to 
results fully independent of the unknown phase in the inelastic region.  This
is one of the important strengths of the method advocated in this work.

For the calculation of the integral defined in (\ref{eq:L2}) we
have used the BaBar data \cite{BABAR} 
from $\tin$ up to $\sqrt{t}=3\, \gev$,  continued with  a constant value for the modulus 
in the range $3\, \gev \leq \sqrt{t} \leq 20 \gev$,  smoothly connected with a $1/t$ decrease above 20 GeV. 
As discussed in \cite{Ananthanarayan:2012tn} these assumptions are expected to overestimate the value of $I$. 
In view of the monotony property mentioned at the end of Sec. \ref{sec:method}, this provides a guarantee 
of robust and conservative bounds.

We have worked in our analysis with weights of the type (\ref{eq:rhogeneric0}). 
As shown in \cite{Ananthanarayan:2012tn}, for weigths decreasing as $1/\sqrt{t}$ or 
faster, the contribution of the asymptotic regime $\sqrt{t}>20\,\gev$ is negligible. 
We have tested a large class of expressions, 
and finally settle down to two suitable weights: $\rho(t)=1/t$ and $\rho(t)=\sqrt{t}/(t+3)$. 
The values of $I$ for  the corresponding choices of the parameters $b, c$ and $Q^2$ 
are given in Table \ref{table:Ia}, 
where the uncertainties are due to the BaBar experimental errors.  

\begin{table}
\begin{center}
\caption{Values of the integral $I$ defined in (\ref{eq:L2}) corresponding  to the weight 
functions (\ref{eq:rhogeneric0}) adopted in our analysis. Note that all energies are in GeV.}\vspace{0.1cm}
\label{table:Ia}
\renewcommand{\tabcolsep}{1.5pc} 
\renewcommand{\arraystretch}{1.1} 
\begin{tabular}{cccc}\hline
$b$  	& $c$ 	& $Q^2$	& $I$ \\\hline 
0 	& 1	& 0		& $0.578 \pm  0.022$   \\
1/2 	& 1	& 3		& $0.246 \pm  0.011$\\ 
\noalign{\smallskip}\hline
\end{tabular}
\end{center}
\end{table}

As discussed in Sec. \ref{sec:intro}, we use as input also some values inside the analyticity domain, as specified in (\ref{eq:taylor}) and (\ref{eq:spacelike}).   
For the charge radius we have adopted the range given in \cite{Colangelo:2004,Masjuan}
\beq\label{eq:r2}
  \la r_\pi^2 \ra =0.43 \pm 0.01 \,\mbox{fm}^2,
\eeq
while for the spacelike data we have taken the values 
\bea\label{eq:Huber}	
F(-1.60\,\gev^2)= 0.243 \pm  0.012_{-0.008}^{+0.019}\,, \nonumber \\ 
F(-2.45\, \gev^2)=  0.167 \pm 0.010_{-0.007}^{+0.013}\,,
\eea
given by the most precise recent experiments \cite{Horn, Huber}.

\begin{figure}[htb]
\vspace{0.35cm}
\includegraphics[width = 7.cm]{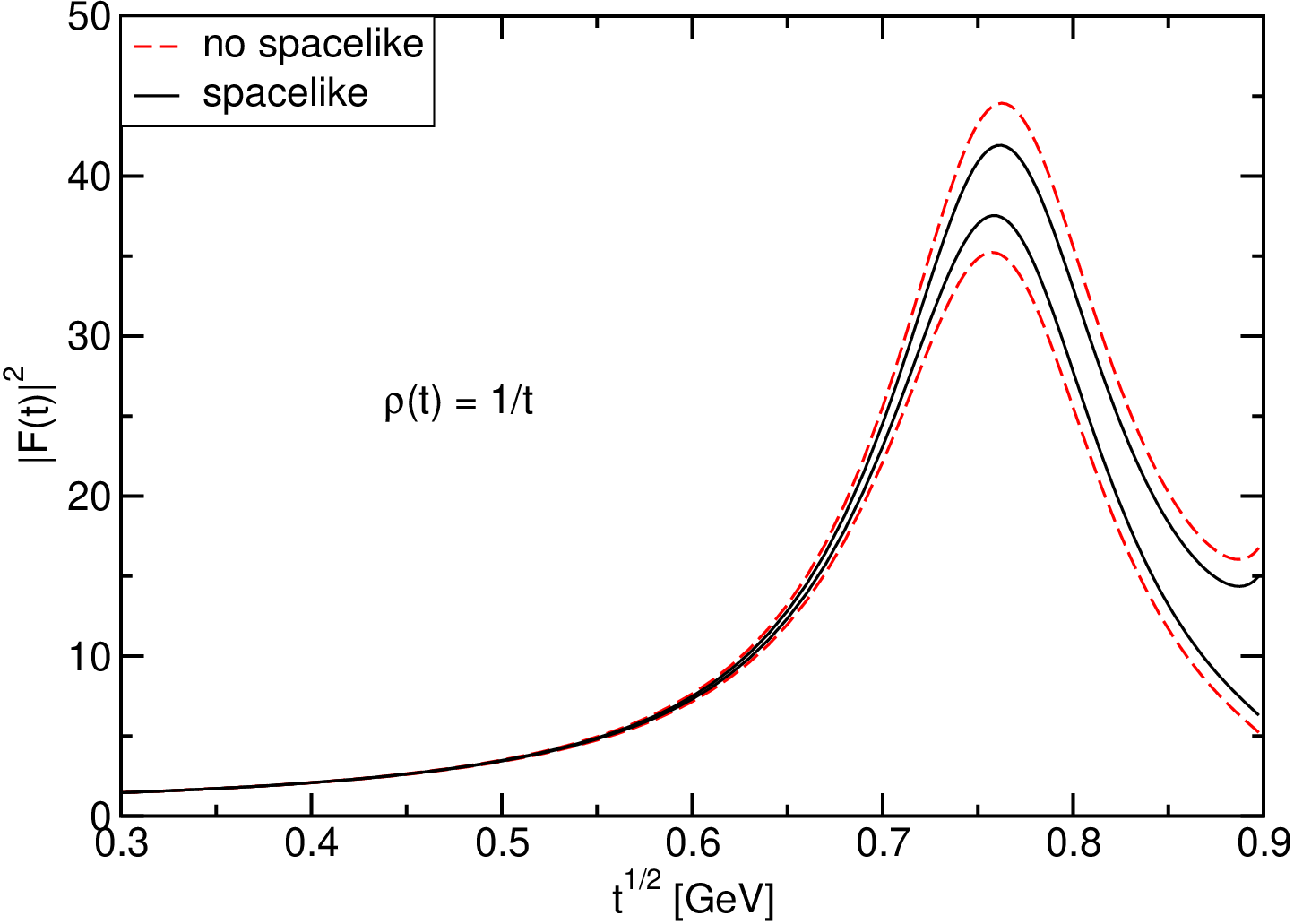}
\caption{Upper and lower bounds on $|F(t)|^2$ below 0.917 GeV for central values of the input, obtained with the weight $\rho(t)=1/t$. Dashed lines: bounds obtained without spacelike input; solid black lines: bounds obtained with spacelike input.}
\label{fig:fig1a}
\end{figure}

\begin{figure}[htb]
\vspace{0.35cm}
 \includegraphics[width = 7.cm]{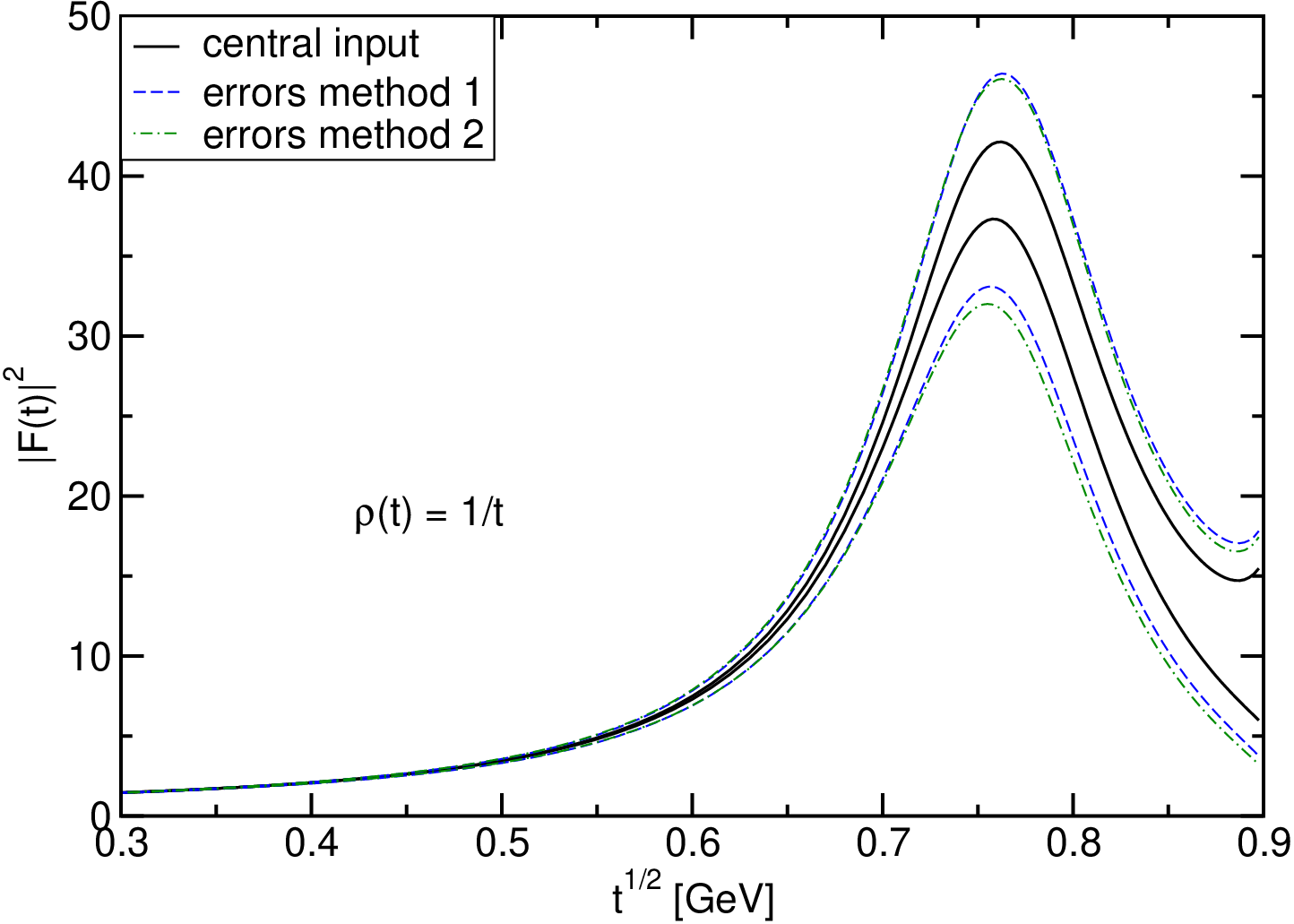}
\caption{Comparison of the two methods of including the uncertainties of the input discussed in the text.}
\label{fig:fig3}
\end{figure}

\begin{figure}[htb]
\vspace{0.35cm}
 \includegraphics[width = 7.cm]{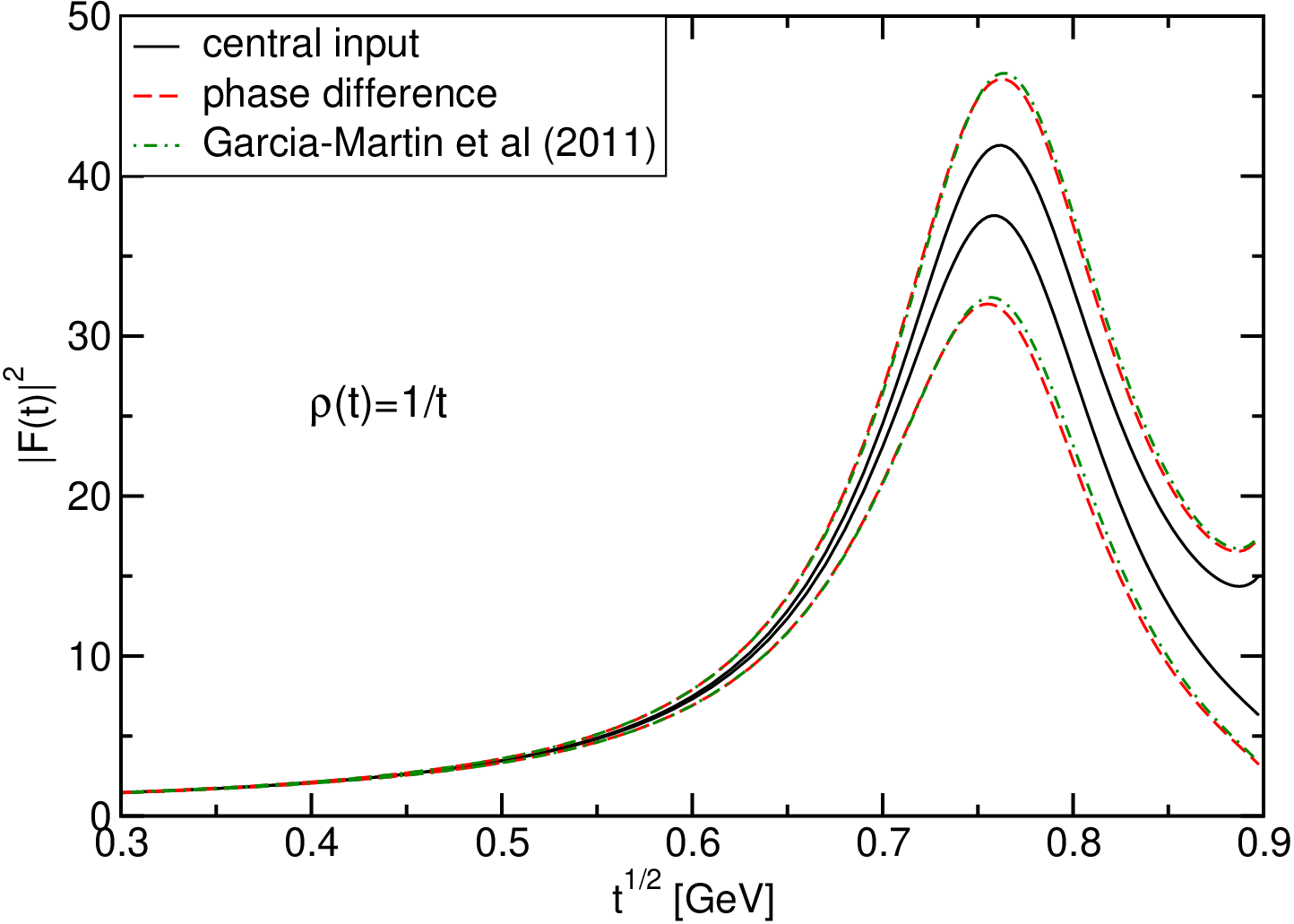}
\caption{Comparison of the two methods of accounting for the errors on the phase shift $\delta_1^1(t)$ discussed in the text. }
\label{fig:fig4}
\end{figure}

\section{Results}\label{sec:results}

In Fig. \ref{fig:fig1a} we show the upper and lower bounds on $|F(t)|^2$ as function of the  centre of mass 
energy $\sqrt{t}$, on the unitarity cut up to the $\omega\pi$ threshold $\sqrt{\tin}= 0.917\,\gev$, derived with the
formalism presented in Sec. \ref{sec:method}.  The curves are obtained with the central values of all the input quantities described in Sec. \ref{sec:inputs}, in the isospin symmetry limit,  using in (\ref{eq:L2}) the weight $\rho(t) = 1/t$. The value $I$ of the integral for this weight is given in Table \ref{table:Ia}.  At each energy, the allowed values of  $|F(t)|^2$ are situated between the two solid black curves. 

The figure  illustrates also the effect of including a spacelike datum:  the dashed red lines are the weaker bounds obtained without the spacelike information. As discussed in \cite{Ananthanarayan:2012tn}, in order to improve the bounds on $F(t)$ at large $t$ on the spacelike axis, it was convenient to use as input the value of $F(t)$ at a larger $|t|$, {\em i.e.} the second value given in  (\ref{eq:Huber}).
For the bounds on the modulus on the timelike axis the effect is not so clear-cut and depends on the energy, but in general it turns out that the first value in (\ref{eq:Huber}) is useful especially for improving the upper bounds, while the second is more efficient for the lower bounds. Therefore, the black solid lines in Fig. \ref{fig:fig1a} are obtained
   using as input the value of $F(-1.60\,\gev^2)$ for the upper bounds and the value 
$F(-2.45\, \gev^2)$ for the lower bounds, and this choice will be made throughout the subsequent analysis.

The uncertainties of the input are expected to lead to weaker bounds and to enlarge the  allowed range for 
the modulus at each energy. In order to investigate this effect we have used two methods: 
\begin{itemize}
\item 
Method 1 -- vary separately each input inside its error interval with the others kept fixed at their central values, and then combine the corresponding errors on the bounds in quadrature. \\
\item
Method 2 -- vary simultaneously all the input variables inside their allowed intervals set by the quoted errors, and
take the most conservative results, namely the largest upper bounds and the smallest lower bounds 
yielded by a compatible set of input quantities inside these intervals. 
\end{itemize}

\begin{figure}[htb]
\vspace{0.35cm}
 \includegraphics[width = 7.cm]{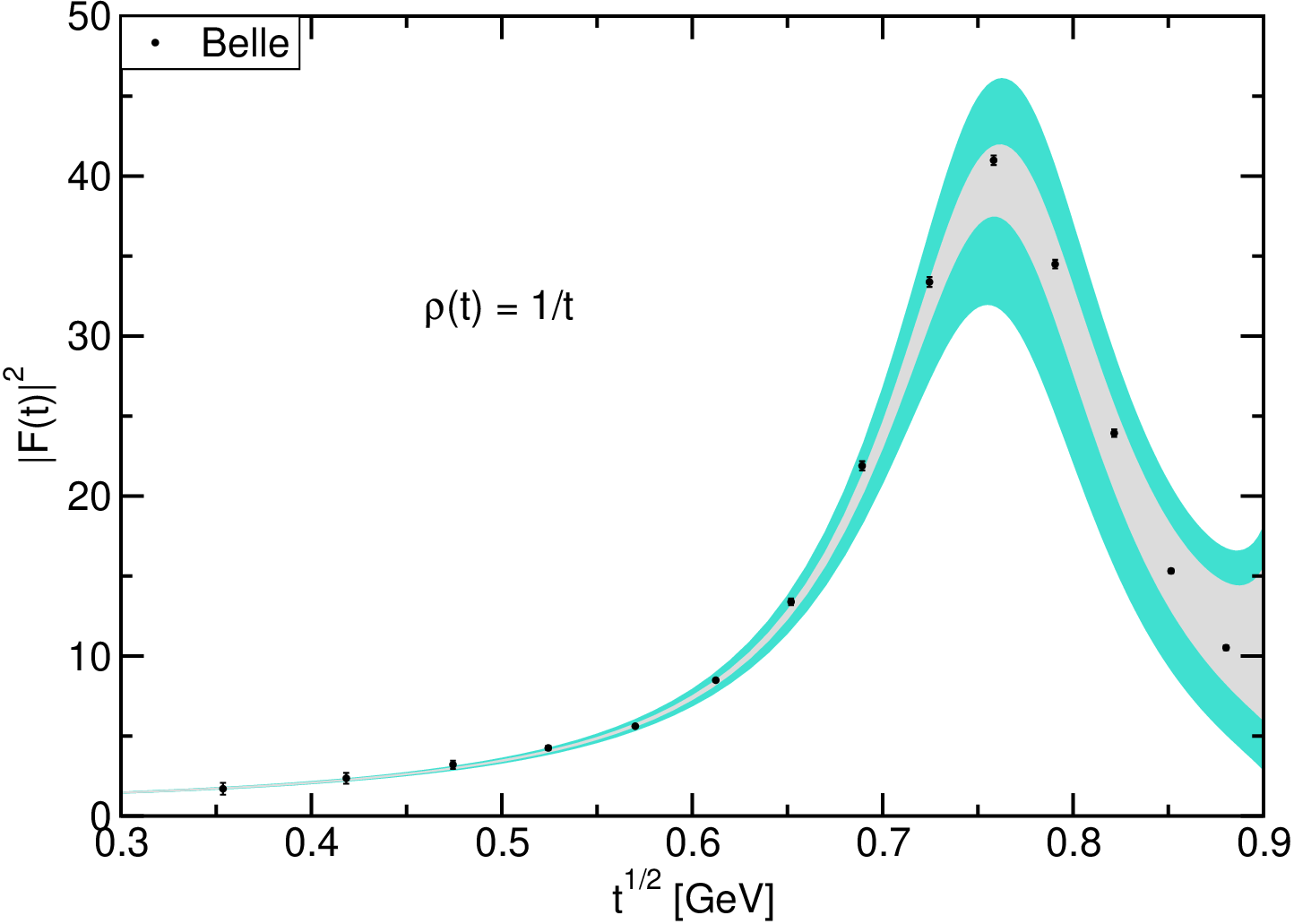}
\caption{Bounds on $|F(t)|^2$ below $\tin$  obtained with the weight $\rho(t) = 1/t$ in the isospin limit. The central grey band is defined by the upper and
the lower bounds obtained with the central values of the input quantities,
the extended cyan bands are obtained by varying all the
inputs within the error intervals. Also shown are the Belle 
data \cite{Fujikawa:2008ma}.}
\label{fig:fig5}
\end{figure}

\begin{figure}[htb]
\vspace{0.35cm}
 \includegraphics[width = 7.cm]{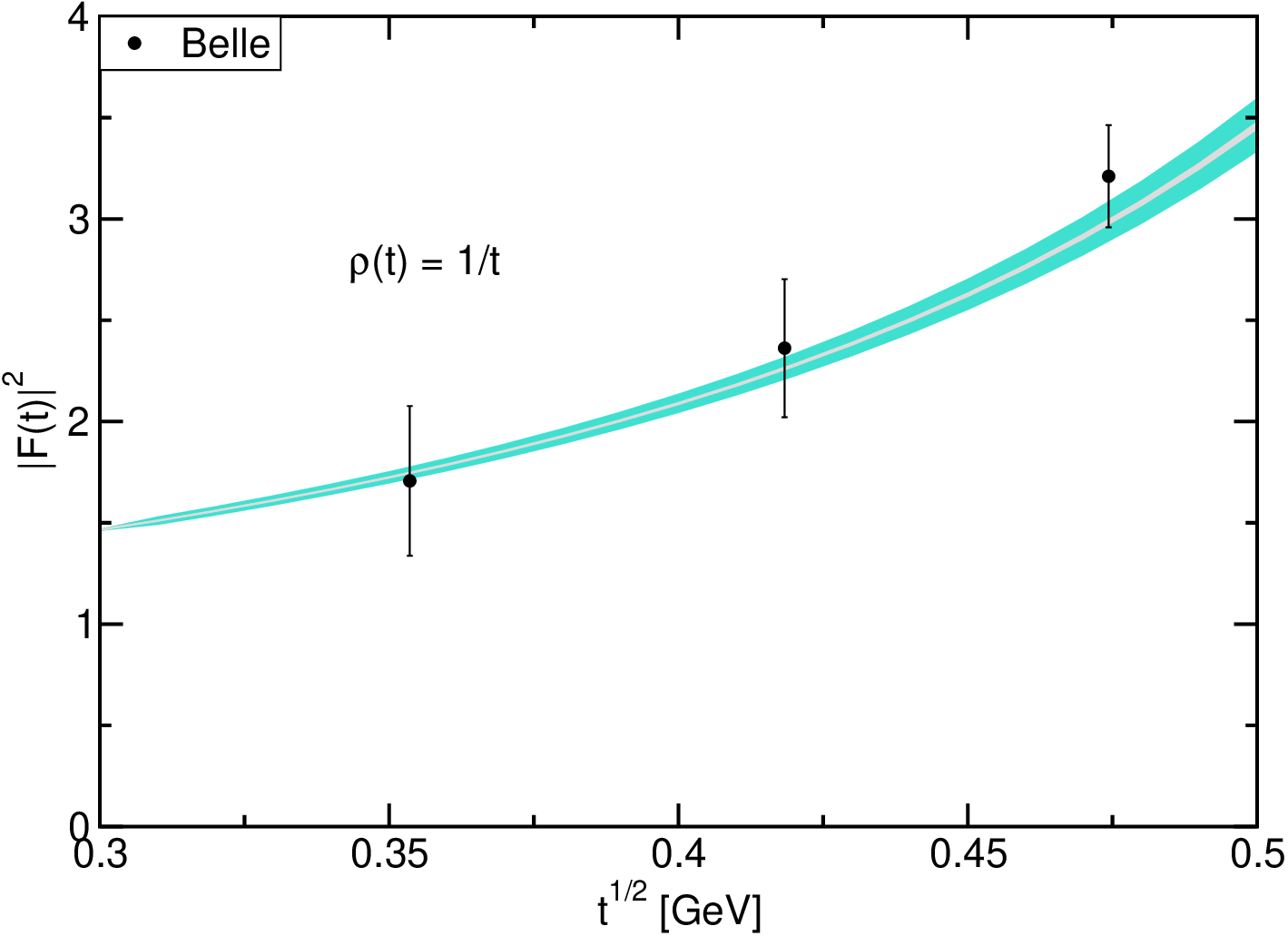}
\caption{Low-energy profile of the bounds obtained with the weight $\rho(t) = 1/t$ in the isospin limit.  For details see caption to Fig. \ref{fig:fig5}.}
\label{fig:fig6}
\end{figure}

We have found that these two methods give similar results, as illustrated in Fig. \ref{fig:fig3}, where 
the bounds are obtained with the weight $\rho(t) = 1/t$ in the isospin limit. The solid black lines are the upper and lower bounds obtained using the central values
of the inputs, while the blue dashed and the green dashed-dot lines  denote the weaker bounds
obtained by including the errors using the first and the second method, respectively. 

In the rest of the analysis, the enlarged allowed intervals for the modulus due to the errors were obtained with the second method described above, which is actually the only one that allows for a proper inclusion of the correlations between the input values. 

We have devoted special attention to the input phase $\delta(t)$ below $\tin$, as it plays an important 
role in the formalism. We have used the phase-shift $\delta_1^1(t)$ and the uncertainty from  
the CDF solution quoted in \cite{GarciaMartin:2011cn}, presented in Eq. (\ref{eq:delta11}) and below it. 
The uncertainty of this phase is very small, even tiny at small energies, approaching half of a degree 
only at the upper end of the energy interval of interest. As a consequence, the effect of the phase 
uncertainty on the bounds is quite small.  Alternatively, we have estimated the error of the phase as the 
difference between the parametrization  (\ref{eq:delta11}) and the Roy solution recently obtained in 
\cite{Caprini:2011ky}. The results of this analysis are presented in Fig. \ref{fig:fig4}, which shows 
the  bounds obtained with the central input and the weaker bounds obtained by varying all the input 
quantities, using for the phase one of the two methods of estimating the uncertainty discussed above. 
As can be seen, the two methods give practically the same results.

We present in what follows a comparison of the bounds with the most recent experimental data  
from $e^+e^-$ annihilation \cite{BABAR, KLOE1, KLOE2, CMD2:1,CMD2:2} and $\tau$ decays 
\cite{Fujikawa:2008ma}. For the comparison with  $e^+e^-$ data  we included in the calculation of the bounds the isospin breaking  via the correction (\ref{eq:iso}). 

\begin{figure}[htb]
\vspace{0.35cm}
 \includegraphics[width = 7.cm]{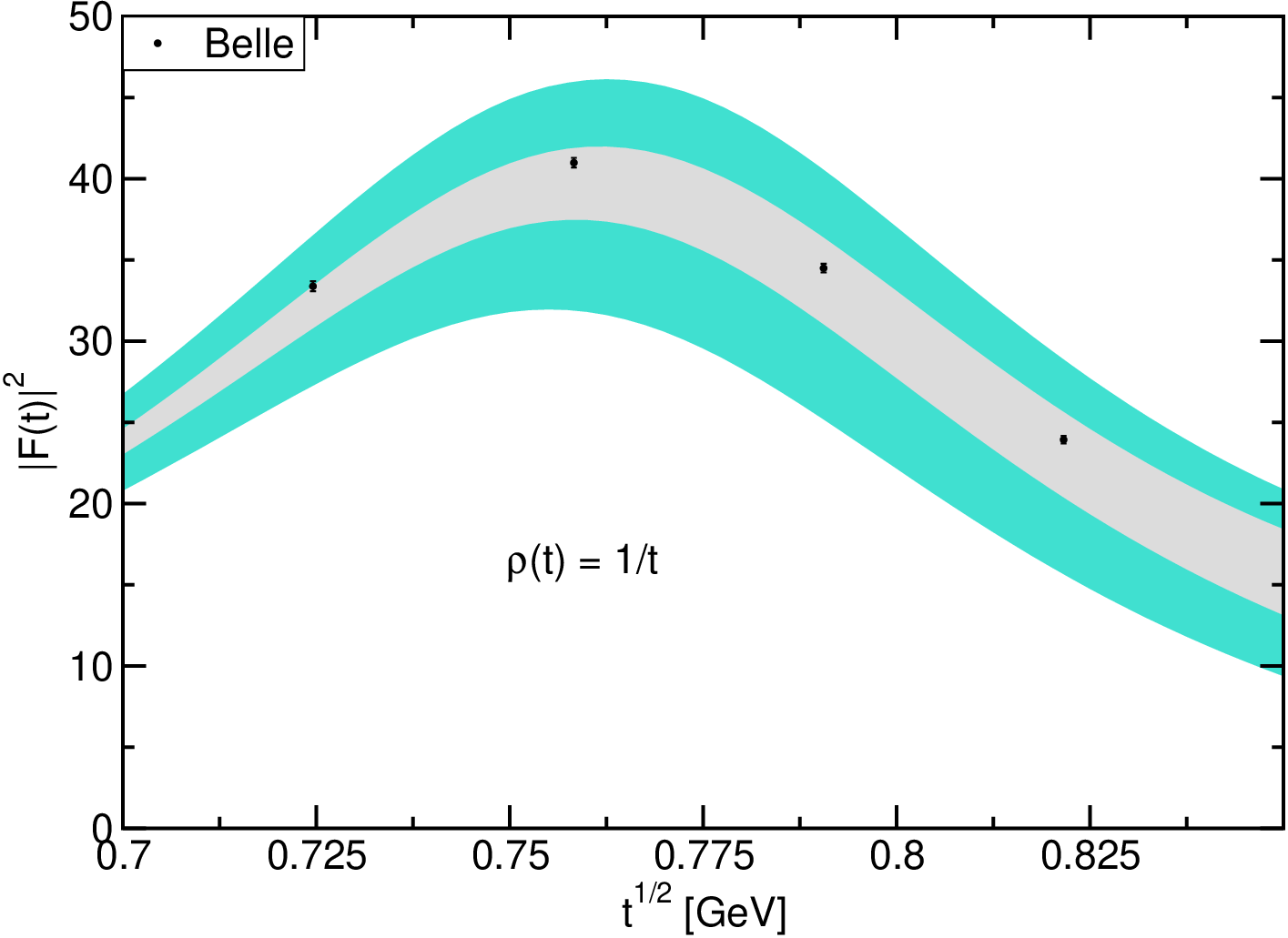}
\caption{Bounds around the resonance region obtained with the weight $\rho(t) = 1/t$ in the isospin limit. The bands are as in Fig. \ref{fig:fig5}.}
\label{fig:fig7}
\end{figure}

\begin{figure}[htb]
\vspace{0.35cm}
 \includegraphics[width = 7.cm]{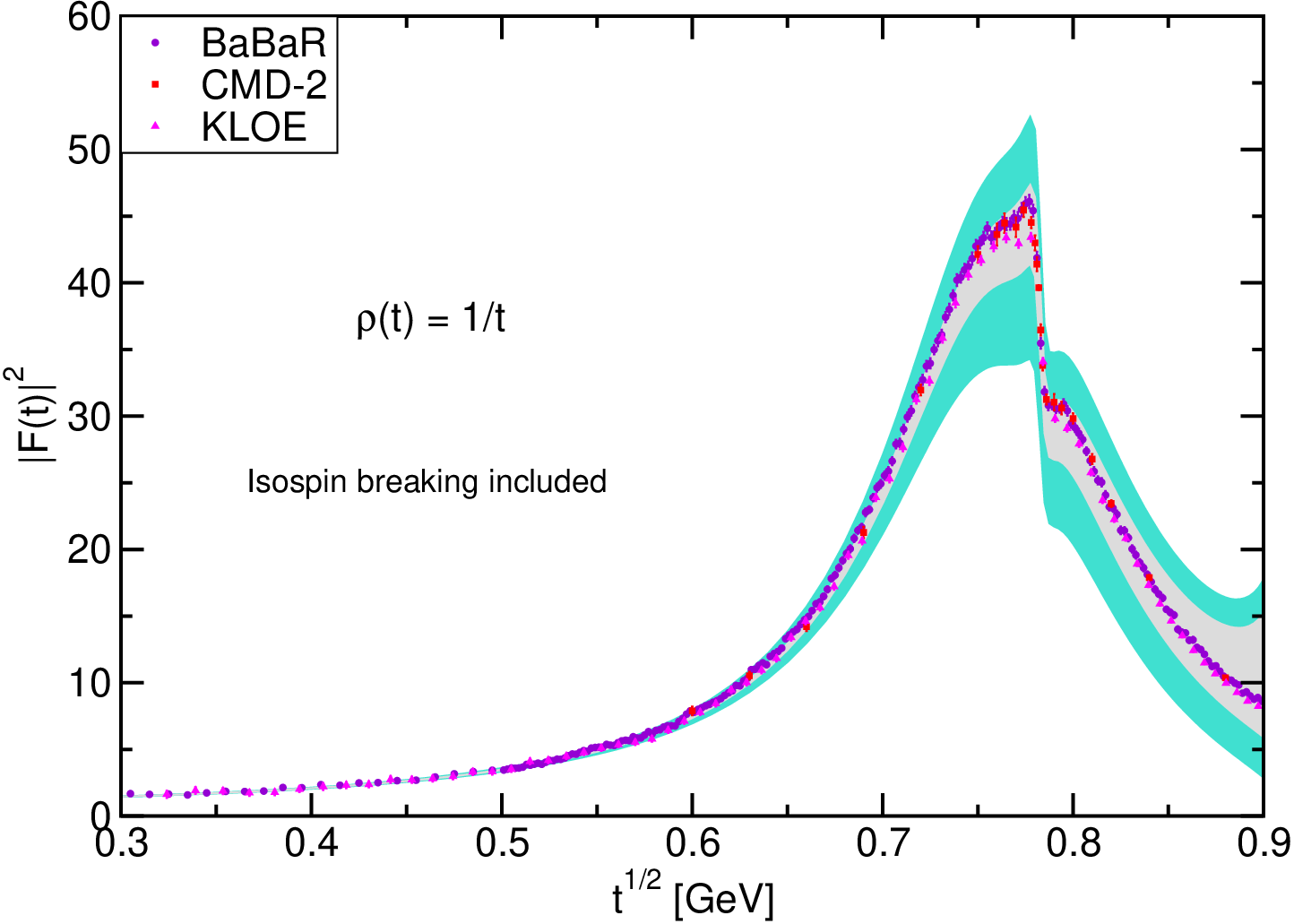}
\caption{Bounds obtained with the weight $\rho(t) = 1/t$ with isospin breaking correction. 
The bands are as in Fig. \ref{fig:fig5}. Also shown are the BaBar \cite{BABAR}, KLOE \cite{KLOE1,KLOE2} and the CMD-2 \cite{CMD2:1,CMD2:2} data.  }
\label{fig:fig8}
\end{figure}

\begin{figure}[htb]
\vspace{0.35cm}
 \includegraphics[width = 7.cm]{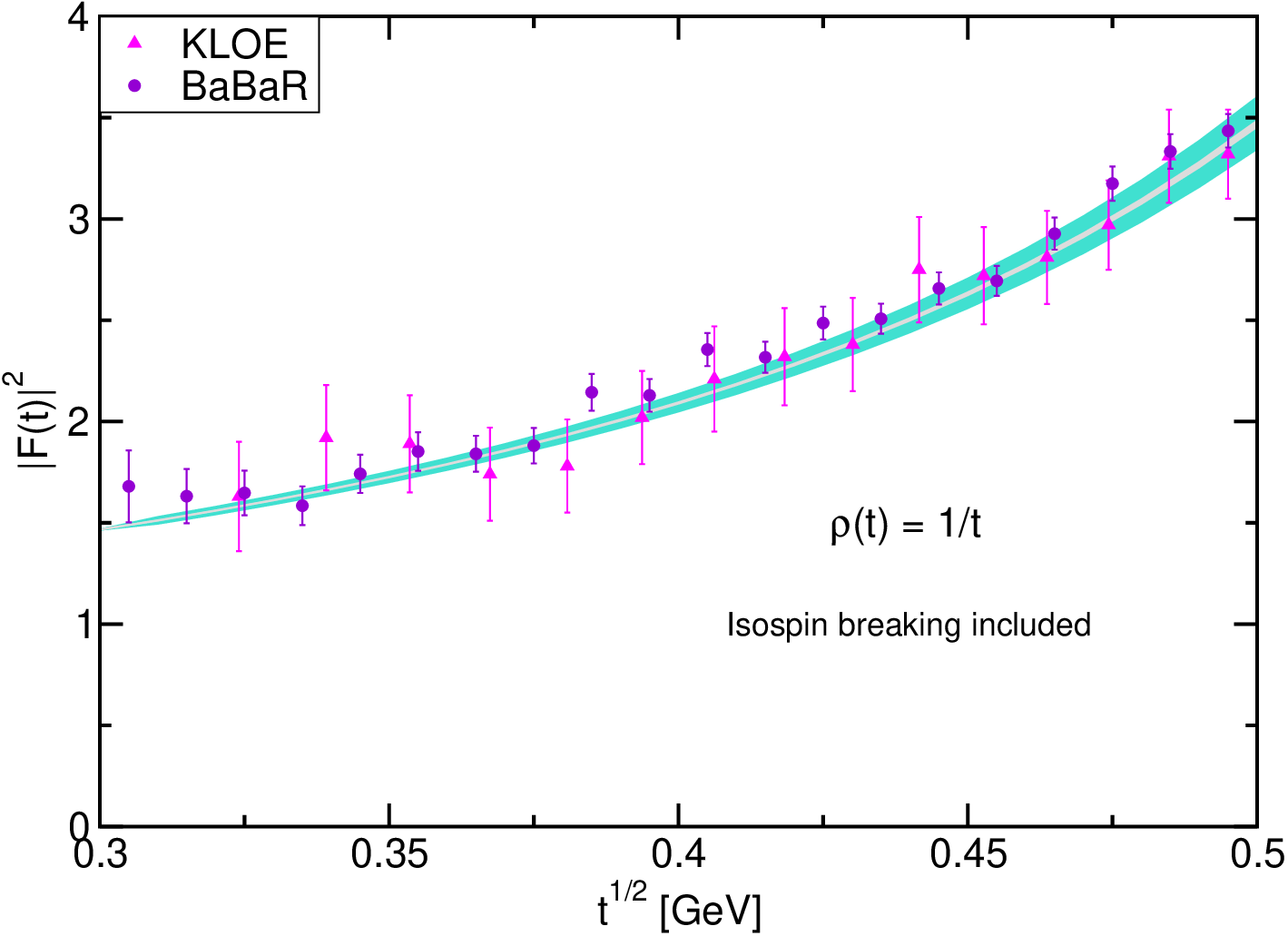}
\caption{Low-energy profile of the bounds obtained with the weight $\rho(t) = 1/t$ with isospin breaking
correction. The bands are as in Fig. \ref{fig:fig5}.}
\label{fig:fig9}
\end{figure}

\begin{figure}[htb]
\vspace{0.35cm}
 \includegraphics[width = 7.cm]{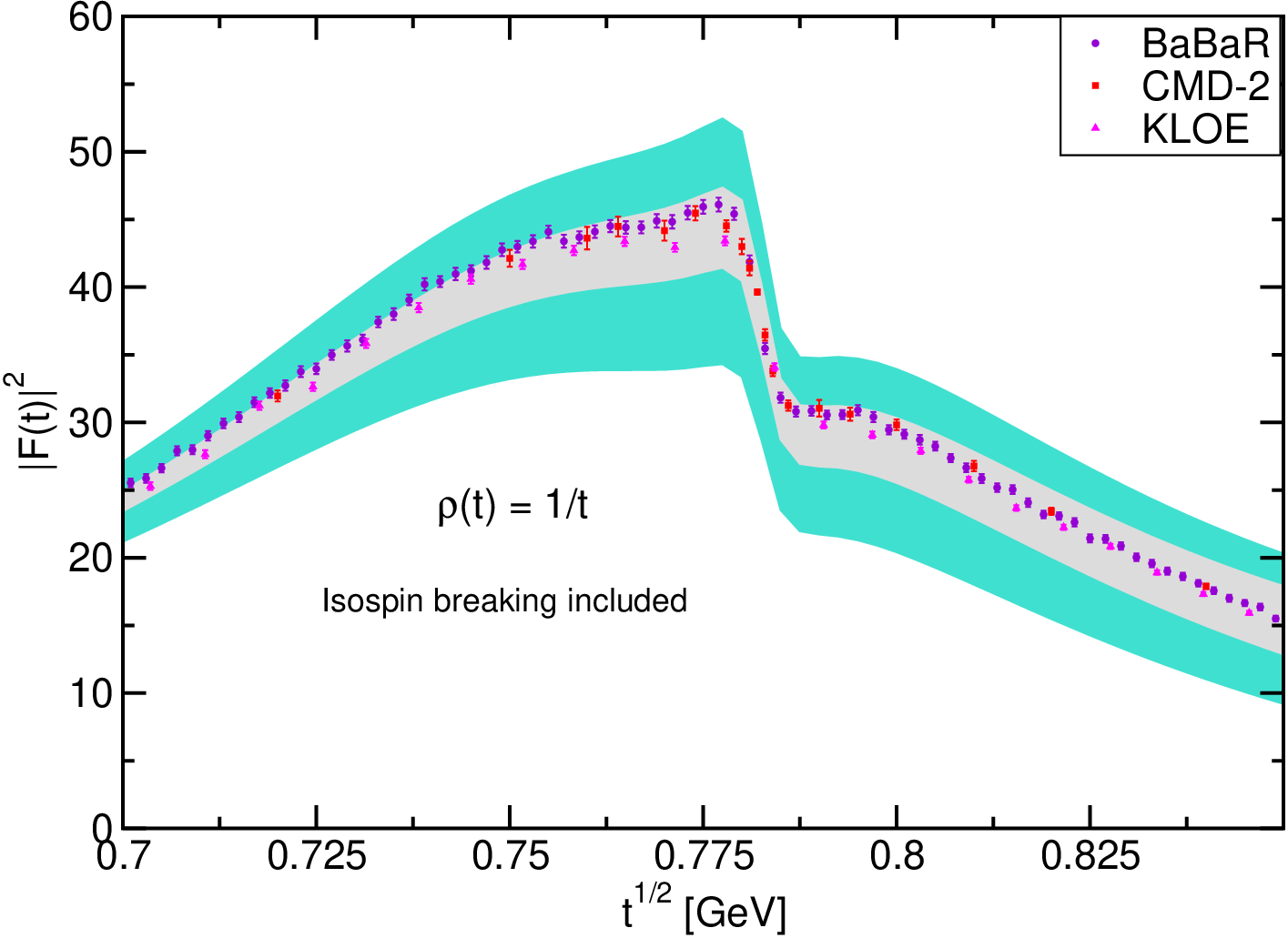}
\caption{Bounds around the resonance region obtained with the weight $\rho(t) = 1/t$ with isospin breaking
correction. The bands are as in Fig. \ref{fig:fig5}.}
\label{fig:fig10}
\end{figure}

The results shown in Figs. \ref{fig:fig5} - \ref{fig:fig10} are obtained with the weight $\rho(t) = 1/t$.  In all of these, 
the central grey bands are defined by the upper and the lower bounds obtained with the central values of input quantities, while 
the extended cyan  bands are obtained by varying all the inputs within the errors as explained above. 
For clarity, we have shown the bounds over the entire region of interest, and also separately in the low-energy  and
the resonance regions.

As shown in Figs. \ref{fig:fig5} and \ref{fig:fig7}, above the $\rho$ peak the Belle 
data \cite{Fujikawa:2008ma}  lie within the grey
band, therefore they are consistent with the bounds obtained with the central input. On the other hand, below the $\rho$ peak  the data are at the upper
edge of the central band. The same features are observed in Figs. \ref{fig:fig8} and \ref{fig:fig10}, where we compare the 
bounds obtained with 
isospin breaking  with the BaBaR \cite{BABAR}, KLOE \cite{KLOE1, KLOE2} and CMD-2 \cite{CMD2:1,CMD2:2}
data. 

\begin{figure}[htb]
\vspace{0.35cm}
 \includegraphics[width = 7.cm]{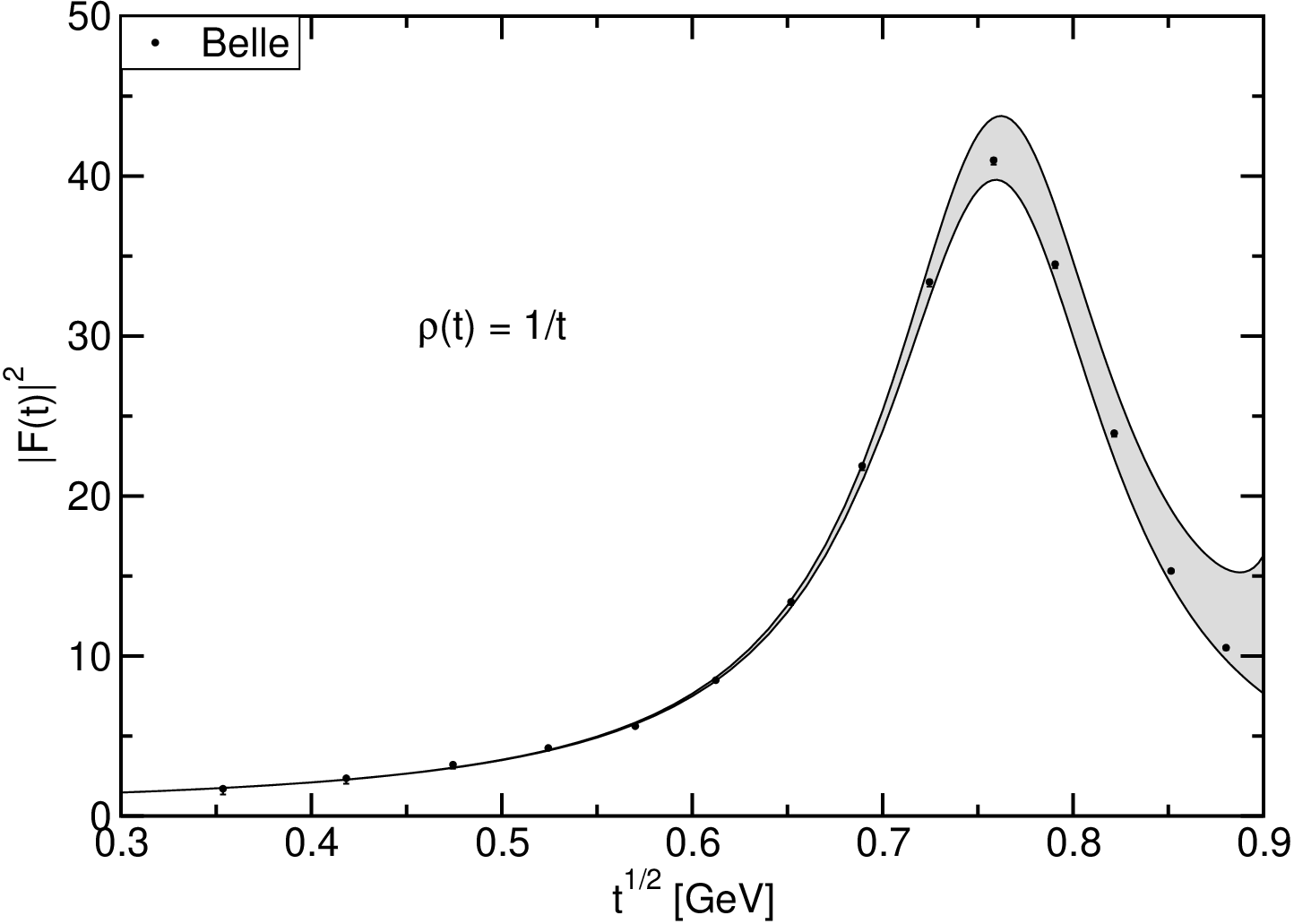}
\caption{Bounds on $|F(t)|^2$ in the isospin limit, obtained with the central values on the input described in the text, 
except for $\la r_\pi^2 \ra =0.435\, \mbox{fm}^2$ instead of the central value in (\ref{eq:r2}). Also shown are the Belle 
data \cite{Fujikawa:2008ma}. }
\label{fig:figrad}
\end{figure}

\begin{figure}[htb]
\vspace{0.45cm}
 \includegraphics[width = 7.cm]{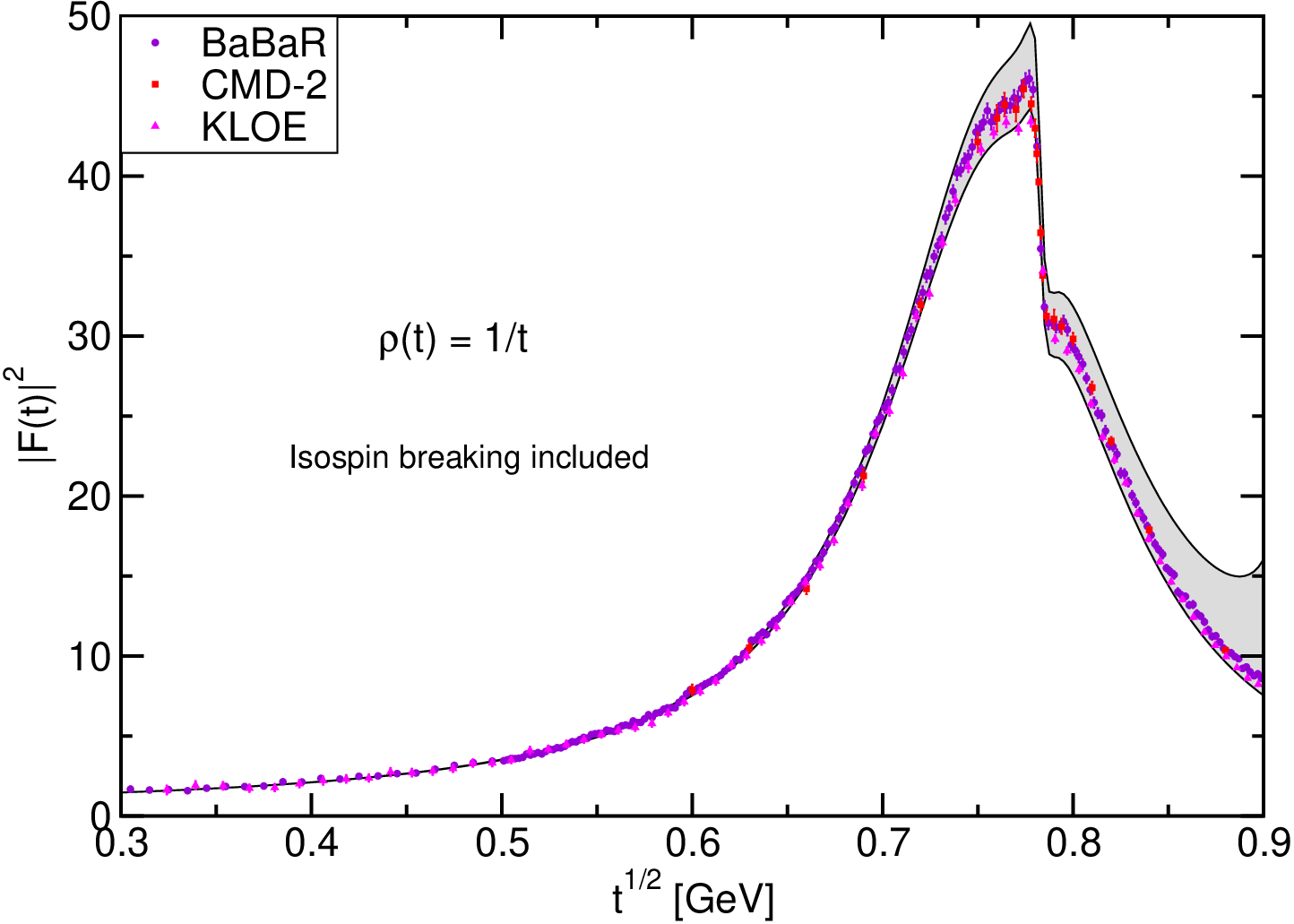}
\caption{As in Fig. \ref{fig:figrad}, also including isospin correction.  Also shown are the BaBar \cite{BABAR}, KLOE \cite{KLOE1,KLOE2} and the CMD-2 \cite{CMD2:1,CMD2:2} data.}
\label{fig:figradiso}
\end{figure}

  The behaviour below the $\rho$ peak deserves attention, because it could   either indicate a problem in the data on the modulus, or suggest that our central input is not optimal. 
We first note that the bounds are quite sensitive to the value of $\la r_\pi^2 \ra$ used as input. For illustration, in Figs. \ref{fig:figrad} and \ref{fig:figradiso} we show the bounds calculated with the central input used above, except for the choice $\la r_\pi^2 \ra =0.435 \,\mbox{fm}^2$ instead of the central value quoted in (\ref{eq:r2}).  This slightly higher value of the charge radius  shifted upwards the bounds by a sizable amount in the resonance region. As a consequence, the data below the $\rho$ peak are  inside the allowed central grey band, but  the data on the right side of the resonance fall now near the lower edge of the allowed region. Therefore, a very good consistency along the whole energy region is not achieved. 
With an even higher central charge radius, like  $\la r_\pi^2 \ra =0.44 \,\mbox{fm}^2$, the lower bound at the left side of the resonance is situated above the data, so larger values of the charge radius lead to an even larger inconsistency. Thus, the variation allowed for the   charge radius is quite limited and cannot fully resolve the slight tension between the recent data on the modulus and the allowed inner band below the $\rho$ peak. 
We must, however, emphasize that the data are fully inside the enlarged  
bands, and as a result there are no inconsistencies if we take into account 
the errors on the input.

One may also question the  phase used as input. Indeed, while the $P$-wave phase shift is obtained in  \cite{ACGL,CGL,KPY,GarciaMartin:2011cn,Caprini:2011ky} by solving Roy equations for the $\pi\pi$ amplitude,  some input to these equations is borrowed in fact from  the knowledge of the pion form factor itself. For instance, the value of the $P$-wave phase shift at the matching point of 0.8 GeV  was taken in \cite{ACGL,CGL,Caprini:2011ky} from a Gounaris-Sakurai parametrization of the CLEO data on the form factor \cite{CLEO}. A comparison of this parametrization with the recent data shows that the latter are systematically larger than the older CLEO data (this is true for both the Belle data from $\tau$ decay \cite{Fujikawa:2008ma}, and for the data \cite{BABAR,KLOE1,KLOE2,CMD2:1,CMD2:2} from $e^+e^-\to\pi^+\pi^-$, after applying the isospin correction). This implies that the input at the matching point 0.8 GeV in the Roy equations might be improved with the new data on the form factor. 

In order to test the sensitivity of the  bounds to the variation of the phase,  we have repeated the calculation with the central phase from  \cite{ACGL,Caprini:2011ky} increased (or decreased) by its quoted error (which is, in particular, of 2 deg at 0.8 GeV). The results show that a higher phase leads to bounds shifted upwards, but it turns out that the change in the calculated bounds is not sufficient  to improve the consistency, within the inner band, on the left side of the $\rho$ peak. However, it appears that the bounds are more sensitive to the detailed shape of the input phase than to its overall magnitude. This aspect deserves further study. 

As we remarked above, for cross-checking the consistency of the data 
we must take into account the enlarged band obtained by allowing the 
inputs to vary within their errors. 
We have displayed  the two kinds of bands since they impressively demonstrate
how an improved input could lead to
improved bounds. Furthermore, the inner band, obtained with 
the central values of the input, allows one to check whether a very good 
consistency holds (close central values are considered in general to indicate a 
better agreement  of two determinations than a mere nonzero overlap of the 
intervals allowed by their error bars). 
Strictly speaking therefore, data lying at the edge of the inner band 
do not point at any tension between the data sets, and our results show 
in fact full consistency between the bounds and the new precise data on the 
modulus.

Turning to the low energy 
region, where the experimental errors are large, Figs. \ref{fig:fig6} and   \ref{fig:fig9}  show that the bounds are very 
stringent, both in the isospin limit and with isospin breaking, offering a more precise description of the modulus than experiment.  This is perhaps the most interesting result 
of our analysis. Also, some tension between several 
points, especially from BaBar, and the bounds  may be noted at low energies, despite the rather 
large experimental errors. 

We have  tested also other weights $\rho(t)$ of the type given in  (\ref{eq:rhogeneric0}). 
Some choices slightly improve the bounds in some energy regions, leading however to weaker bounds in other regions. 
Also, while one weight improves the upper bounds, other weights may lead to better lower bounds. 
In general,  if we limit ourselves to bounds with a suitable decrease at large $t$ such as to ensure a 
reliable calculation of the integral (\ref{eq:L2}), the differences in the resulting bounds are not very large. 
For illustration, we present below 
the results for the weight  $\rho(t) = t^{1/2}/(t+3)$, which leads to the value of the integral  (\ref{eq:L2})
given in Table \ref{table:Ia}.

\begin{figure}[htb]
\vspace{0.35cm}
 \includegraphics[width = 7.cm]{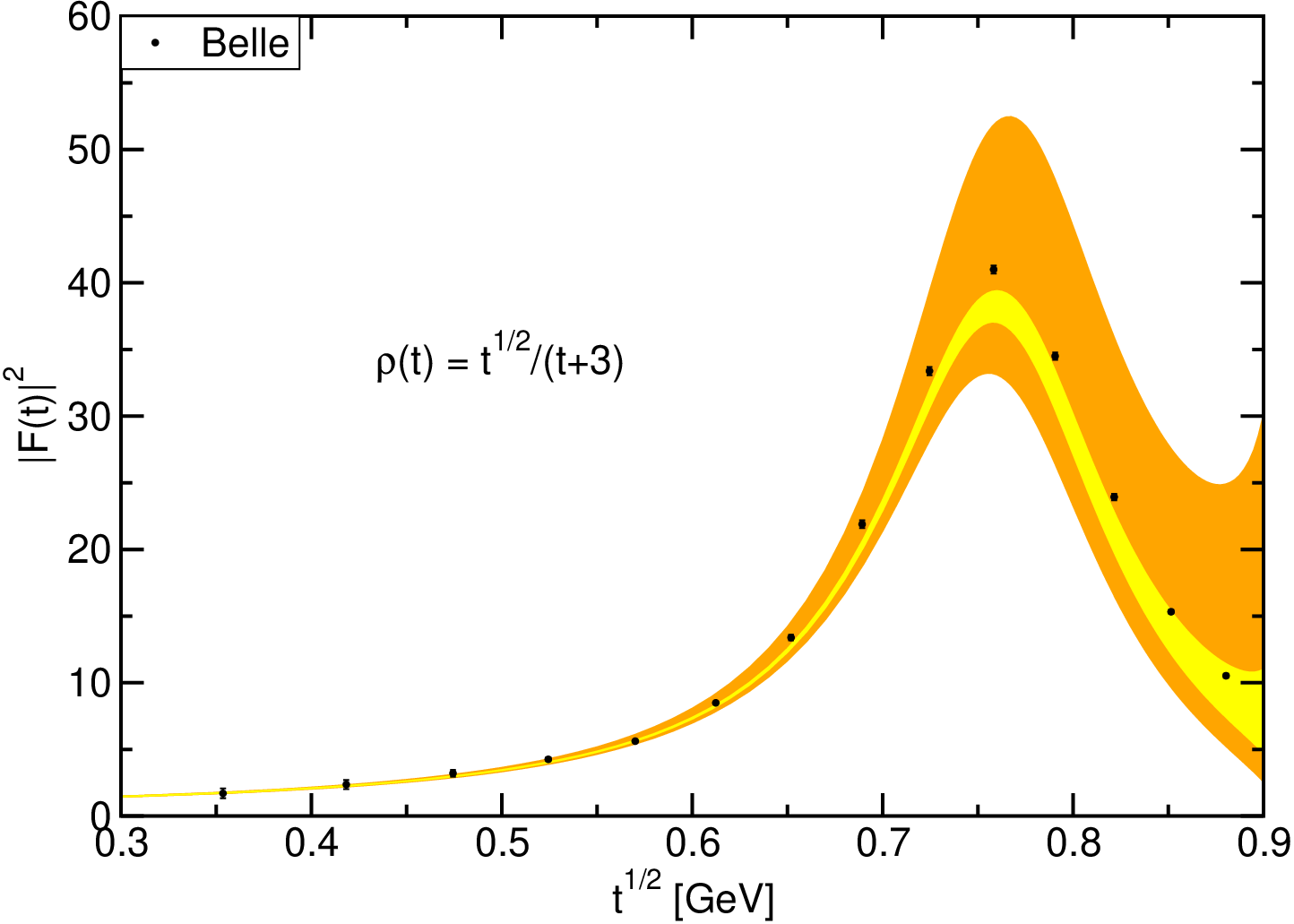}
\caption{Bounds on $|F(t)|^2$ below $\tin$ obtained with the weight $\rho(t) = t^{1/2}/(t+3)$ in the isospin limit.  The central yellow band is defined by the upper and
the lower bounds obtained with the central values of the input quantities,
the extended orange  bands are obtained by varying all the
inputs within the errors. Also shown are the Belle 
data \cite{Fujikawa:2008ma}.}
\label{fig:fig11}
\end{figure}

\begin{figure}[htb]
\vspace{0.45cm}
 \includegraphics[width = 7.cm]{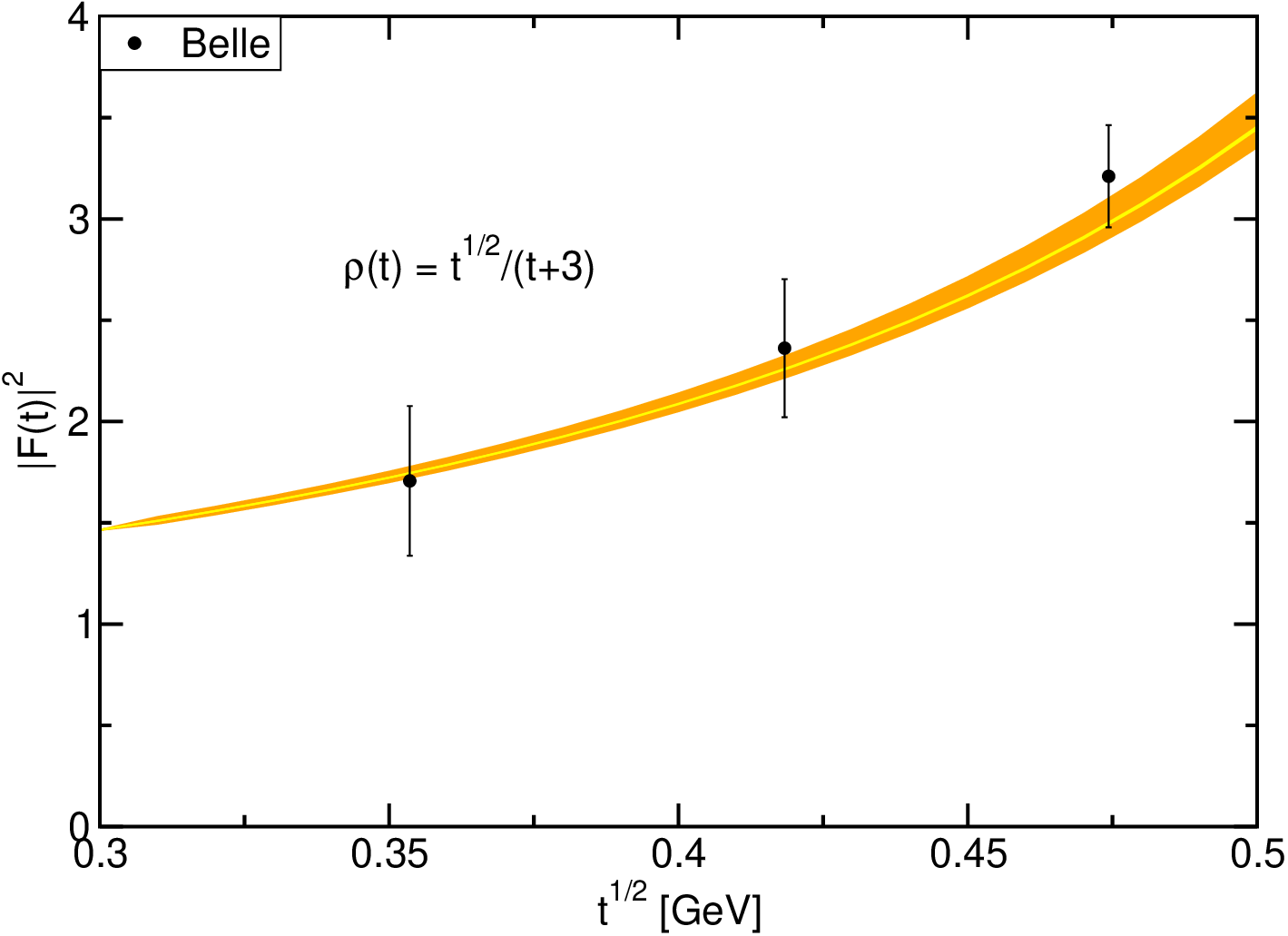}
\caption{Low-energy profile of the bounds obtained with the weight $\rho(t) = t^{1/2}/(t+3)$ in the isospin limit. For details see caption to Fig. \ref{fig:fig11}.}
\label{fig:fig12}
\end{figure}

\begin{figure}[htb]
\vspace{0.35cm}
 \includegraphics[width = 7.cm]{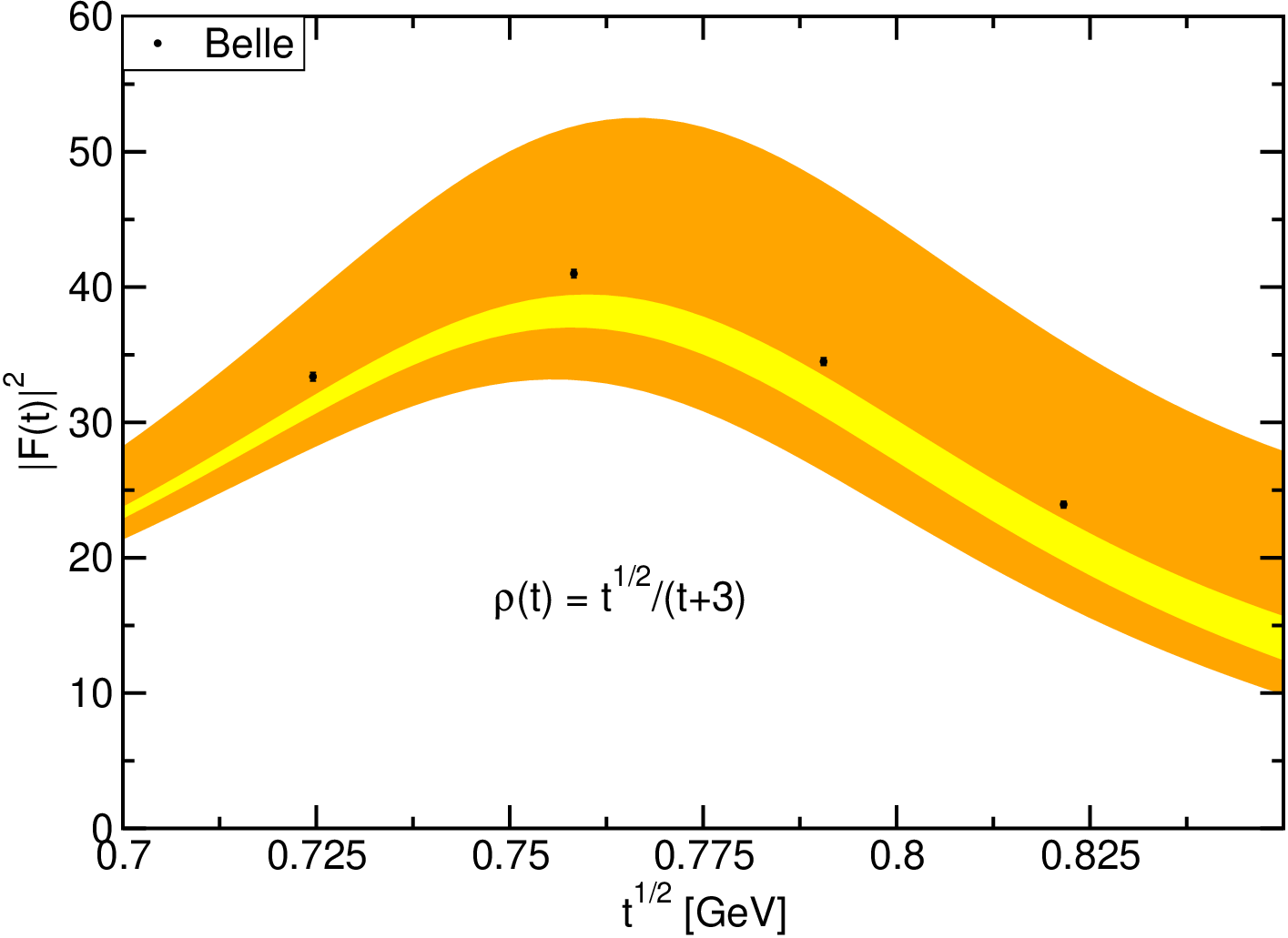}
\caption{Bounds around the resonance region obtained with the weight $\rho(t) = t^{1/2}/(t+3)$ in the isospin limit. The bands are as in
Fig. \ref{fig:fig11}.}
\label{fig:fig13}
\end{figure}

\begin{figure}[htb]
\vspace{0.45cm}
 \includegraphics[width = 7.cm]{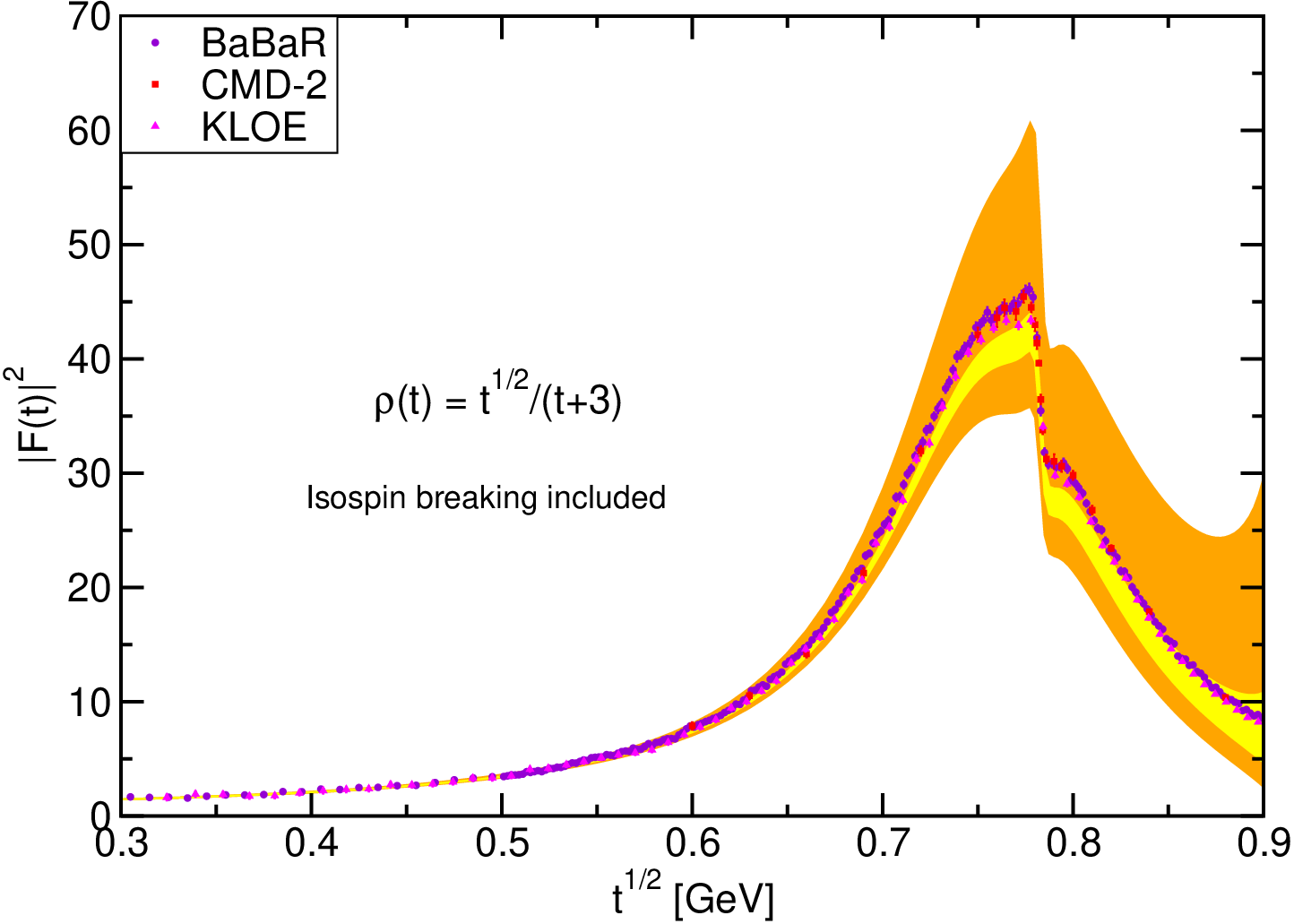}
\caption{Bounds obtained with the weight $\rho(t) = t^{1/2}/(t+3)$ with isospin breaking correction. The bands are as in  Fig. \ref{fig:fig11}.
Also shown are the BaBar \cite{BABAR}, KLOE \cite{KLOE1,KLOE2} and the CMD-2 \cite{CMD2:1,CMD2:2} data. }
\label{fig:fig14}
\end{figure}

\begin{figure}[htb]
\vspace{0.35cm}
 \includegraphics[width = 7.cm]{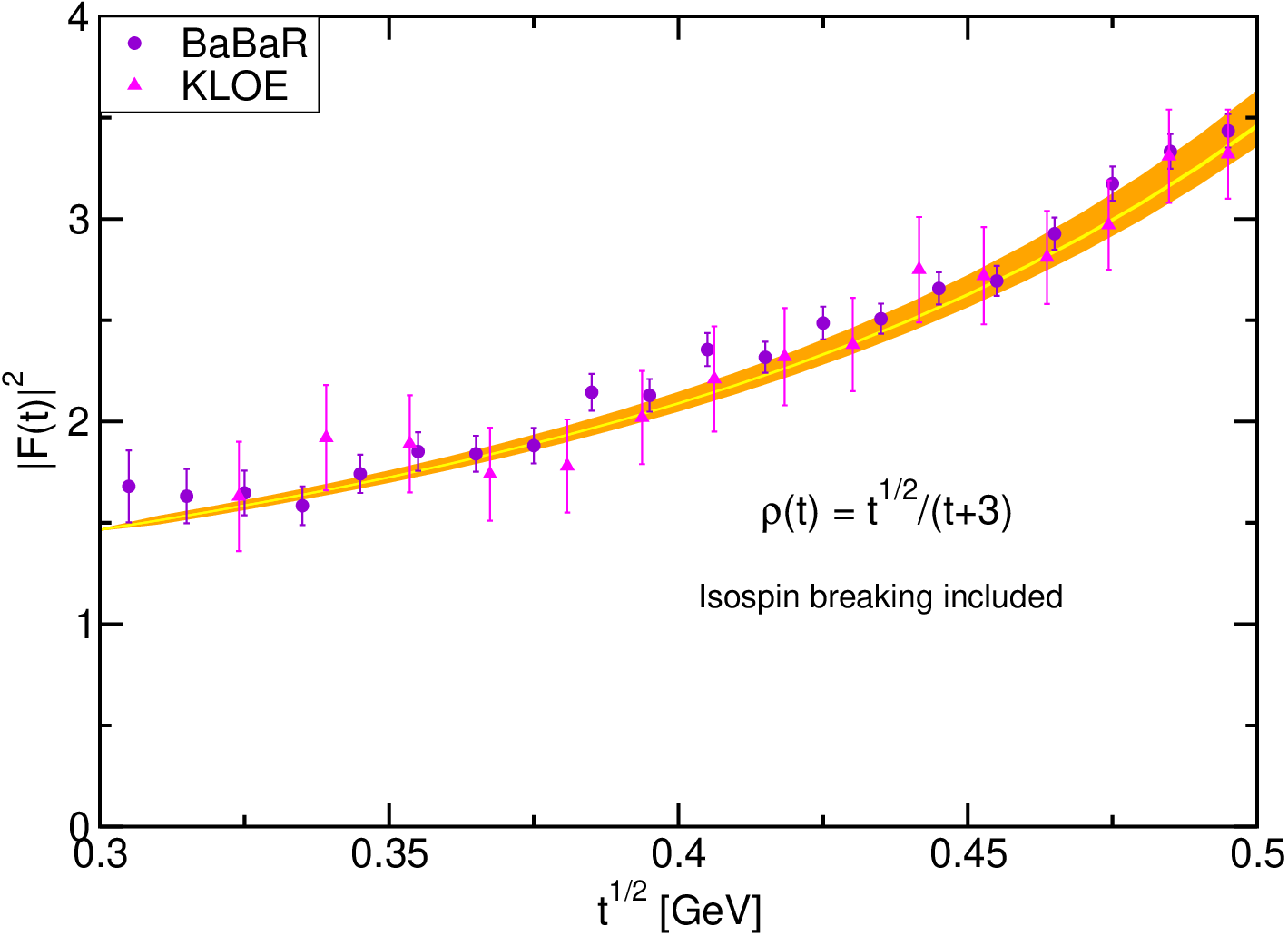}
\caption{Low-energy profile of the bounds obtained with the weight $\rho(t) = t^{1/2}/(t+3)$ with isospin breaking
correction. The bands are as in Fig. \ref{fig:fig11}.}
\label{fig:fig15}
\end{figure}

\begin{figure}[htb]
\vspace{0.4cm}
 \includegraphics[width = 7.cm]{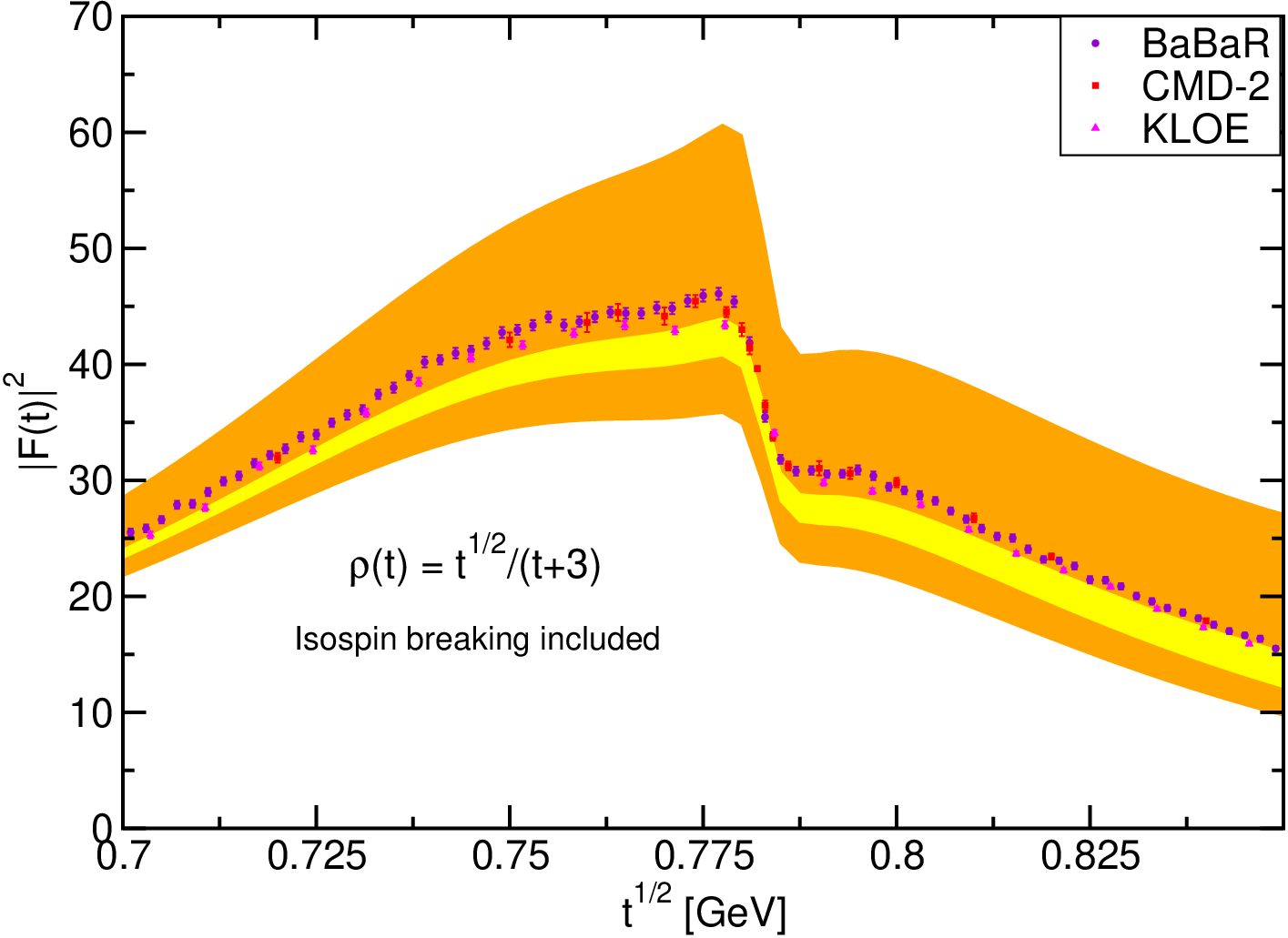}
\caption{Bounds around the resonance region obtained with the weight $\rho(t) = t^{1/2}/(t+3)$ with isospin breaking
correction.  The bands are as in Fig. \ref{fig:fig11}.}
\label{fig:fig16}
\end{figure}

 In Figs. \ref{fig:fig11} - \ref{fig:fig13}  we show the bounds obtained in the isospin limit, 
while Figs. \ref{fig:fig14} - \ref{fig:fig16} are obtained by including the isospin correction (\ref{eq:iso}). The central yellow bands are the allowed ranges for the modulus squared  obtained with the central values of the input quantities, while the orange bands
are obtained by varying all the inputs within the errors. For central values of the input this weight  leads to  upper bounds slightly lower than those obtained with $\rho(t)=1/t$. As a consequence, the experimental 
data on the modulus lie now slightly above the central yellow band, both in the isospin limit and in the symmetry breaking case. Thus, the slight inconsistency remarked above for the bounds obtained with the weight  $\rho(t)=1/t$ is even more pronounced when the bounds are calculated with the  weight $\rho(t) = t^{1/2}/(t+3)$.   However, the upper orange band is now wider and extends to higher values, so this weight finally gives weaker upper bounds than the prior one if the uncertainties are taken into account.  On the other hand, the lower bounds are slightly better in the resonance region and at higher energies than those given by the weight $\rho(t)=1/t$, even when the  uncertainties are taken into account. In the low-energy region, as for the previous weight,  the bounds are very tight.

\begin{figure}[htb]
\vspace{0.35cm}
 \includegraphics[width = 7.cm]{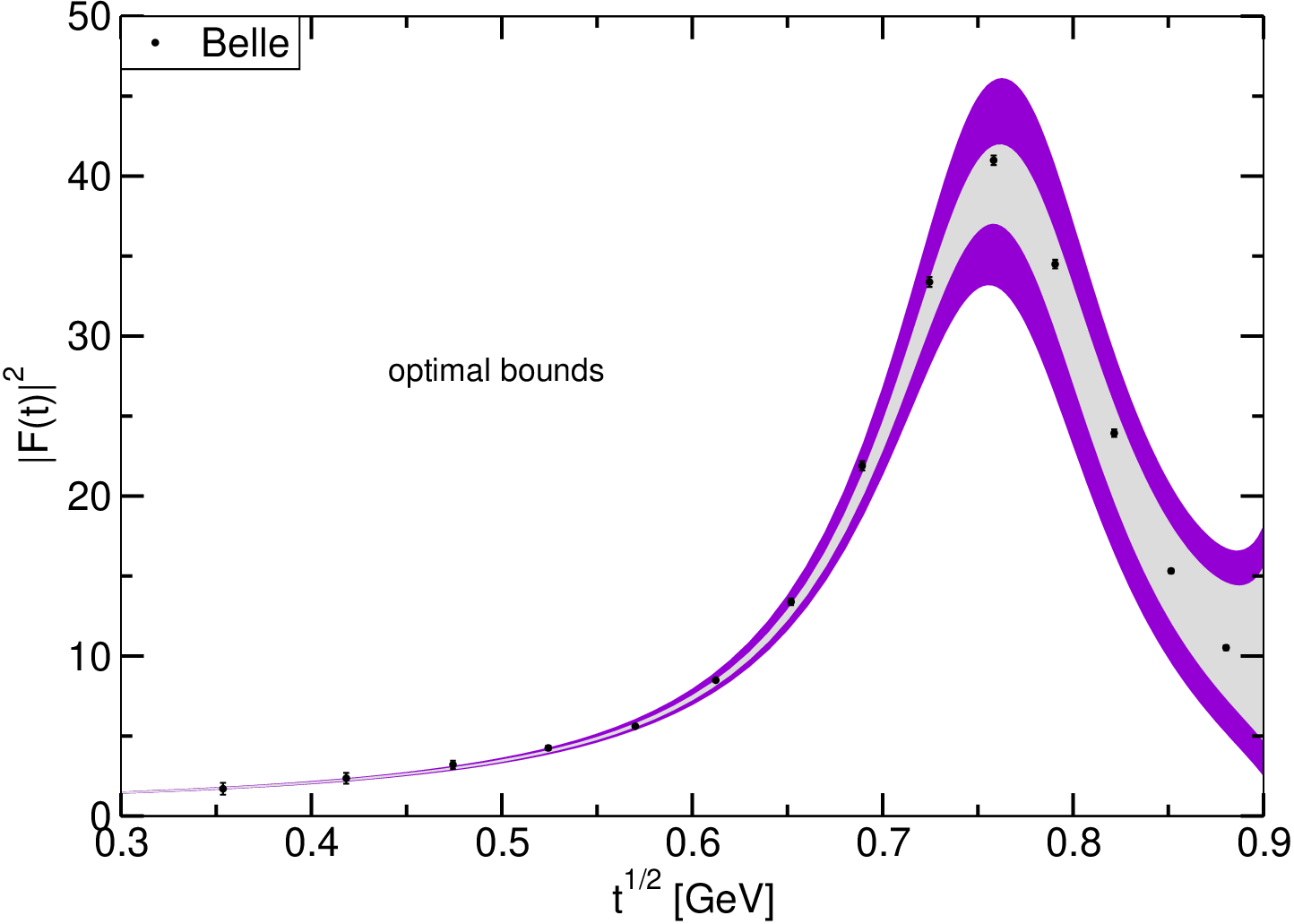}
\caption{Optimal bounds  on $|F(t)|^2$ below $\tin$ in the isospin limit, obtained with the weight $\rho(t) = 1/t$ for the upper bounds and 
$\rho(t) = t^{1/2}/(t+3)$ for the lower bounds.  The central grey band is defined by the upper and
the lower bounds obtained with the central values of the input quantities,
the extended violet bands  are obtained by varying all the
inputs within the errors. Also shown are the Belle 
data \cite{Fujikawa:2008ma}.}
\label{fig:fig17}
\end{figure}

\begin{figure}[htb]
\vspace{0.45cm}
 \includegraphics[width = 7.cm]{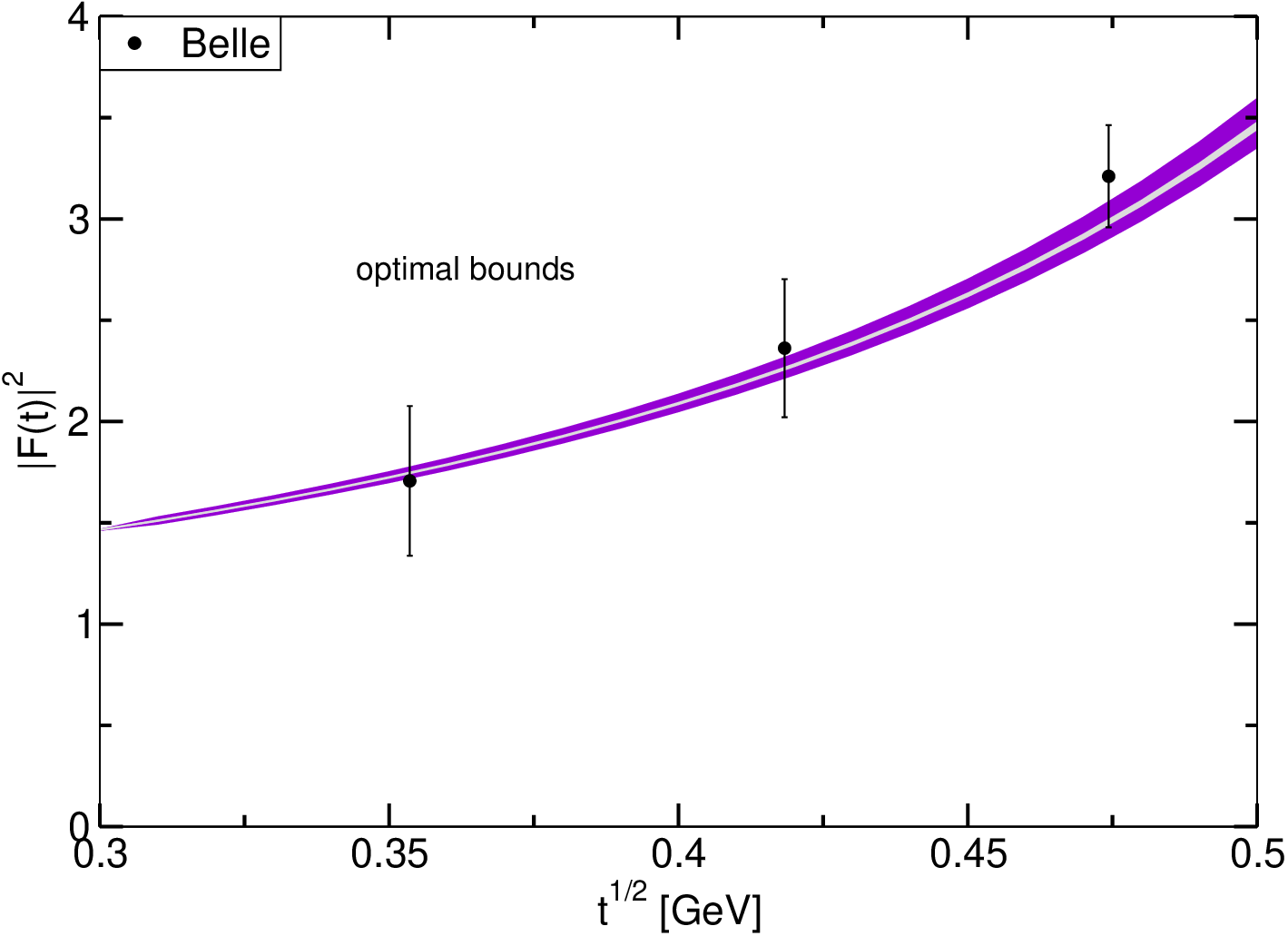}
\caption{Low-energy profile of the optimal bounds in the isospin limit. For details see caption to Fig. \ref{fig:fig17}.}
\label{fig:fig18}
\end{figure}

\begin{figure}[htb]
\vspace{0.35cm}
 \includegraphics[width = 7.cm]{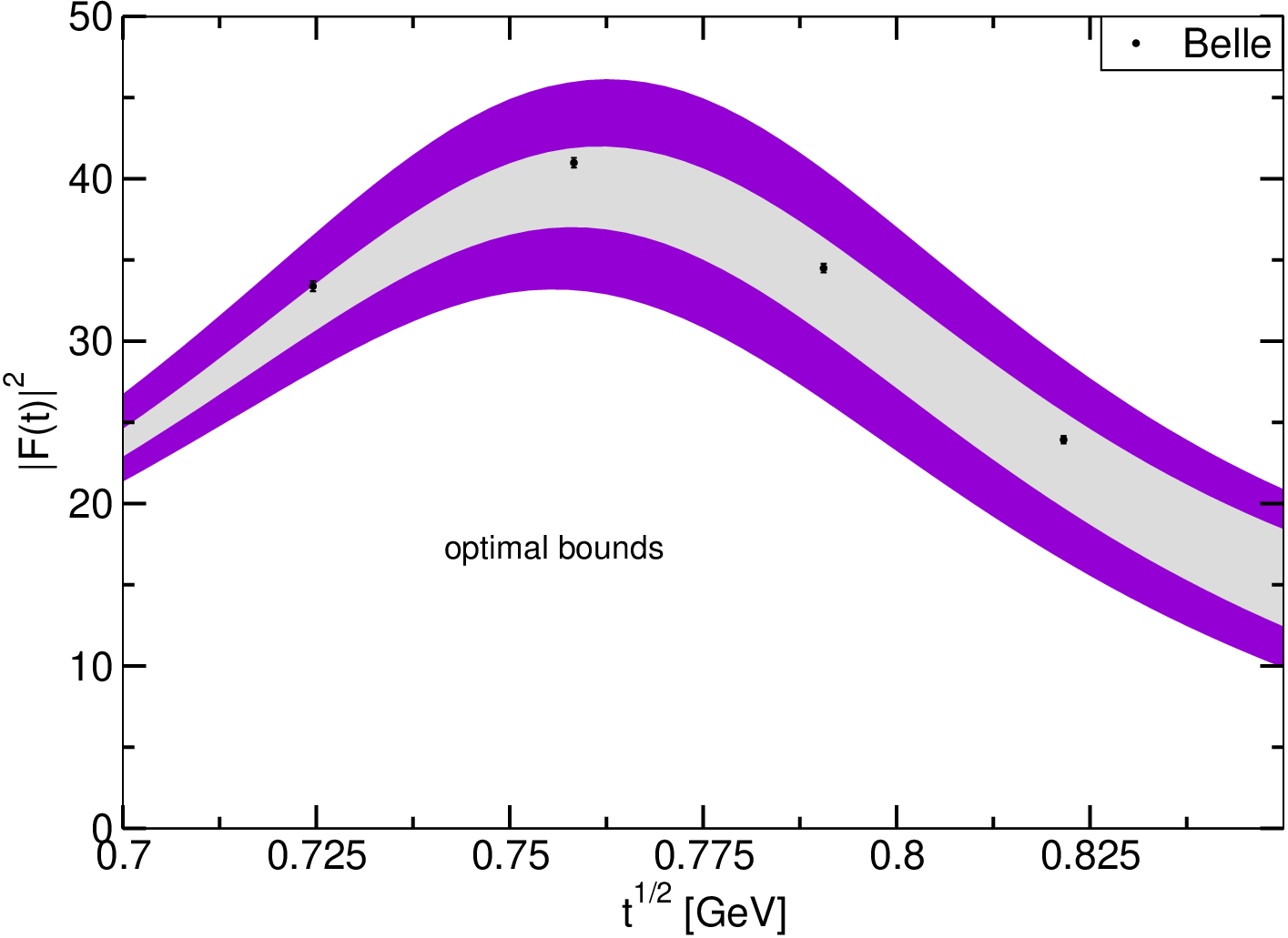}
\caption{Optimal bounds around the resonance region in the isospin limit. The bands are as in Fig. \ref{fig:fig17}.}
\label{fig:fig19}
\end{figure}

\begin{figure}[htb]
\vspace{0.35cm}
 \includegraphics[width = 7.cm]{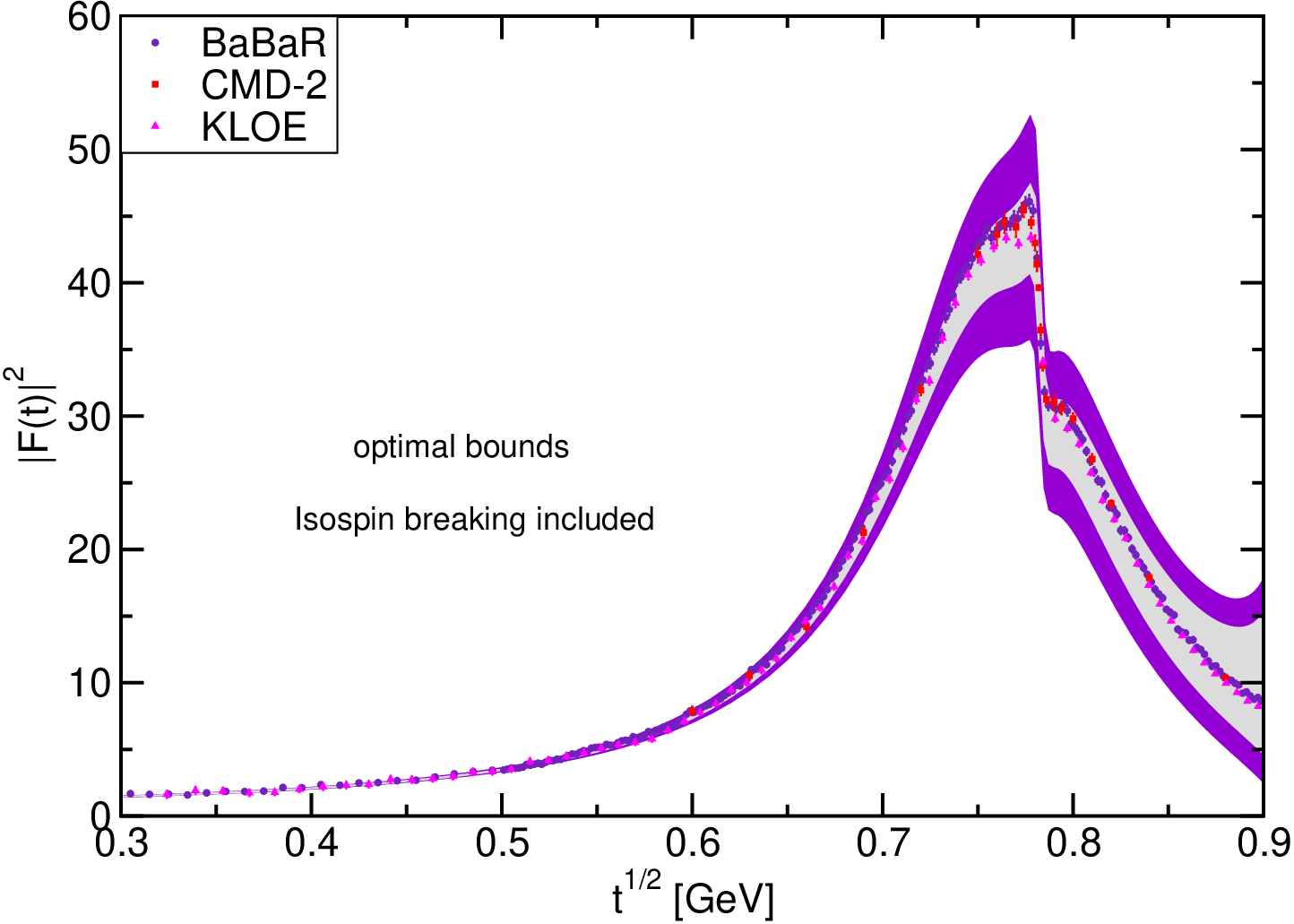}
\caption{Optimal bounds with isospin breaking correction. The bands are as in caption to Fig. \ref{fig:fig17}. 
Also shown are the BaBar \cite{BABAR}, KLOE \cite{KLOE1,KLOE2} and the  CMD-2 \cite{CMD2:1,CMD2:2} data. }
\label{fig:fig20}
\end{figure}

\begin{figure}[htb]
\vspace{0.35cm}
 \includegraphics[width = 7.cm]{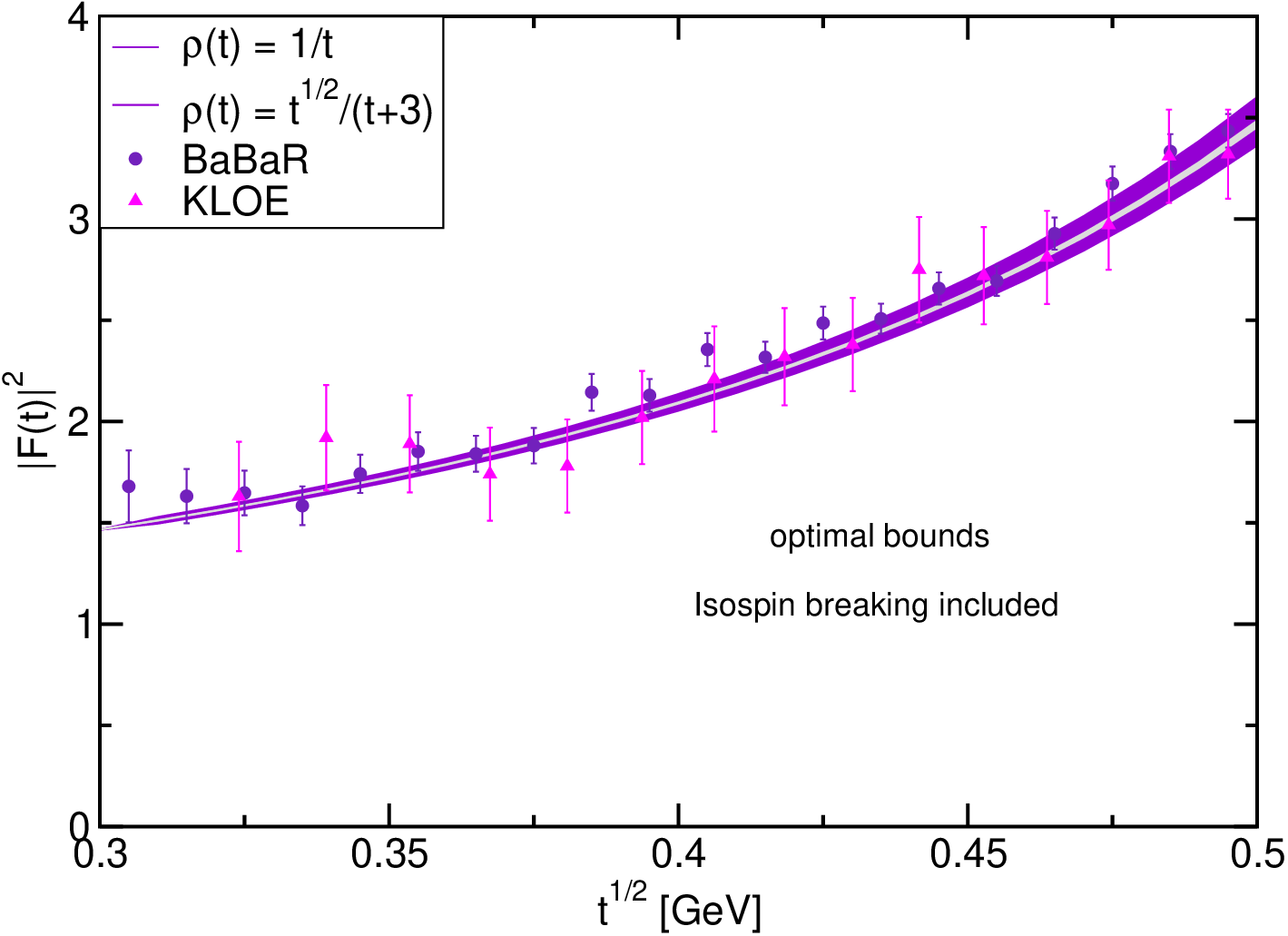}
\caption{Low-energy profile of the optimal bounds with isospin breaking correction. The bands are as in Fig. \ref{fig:fig17}.}
\label{fig:fig21}
\end{figure}

\begin{figure}[htb]
\vspace{0.45cm}
 \includegraphics[width = 7.cm]{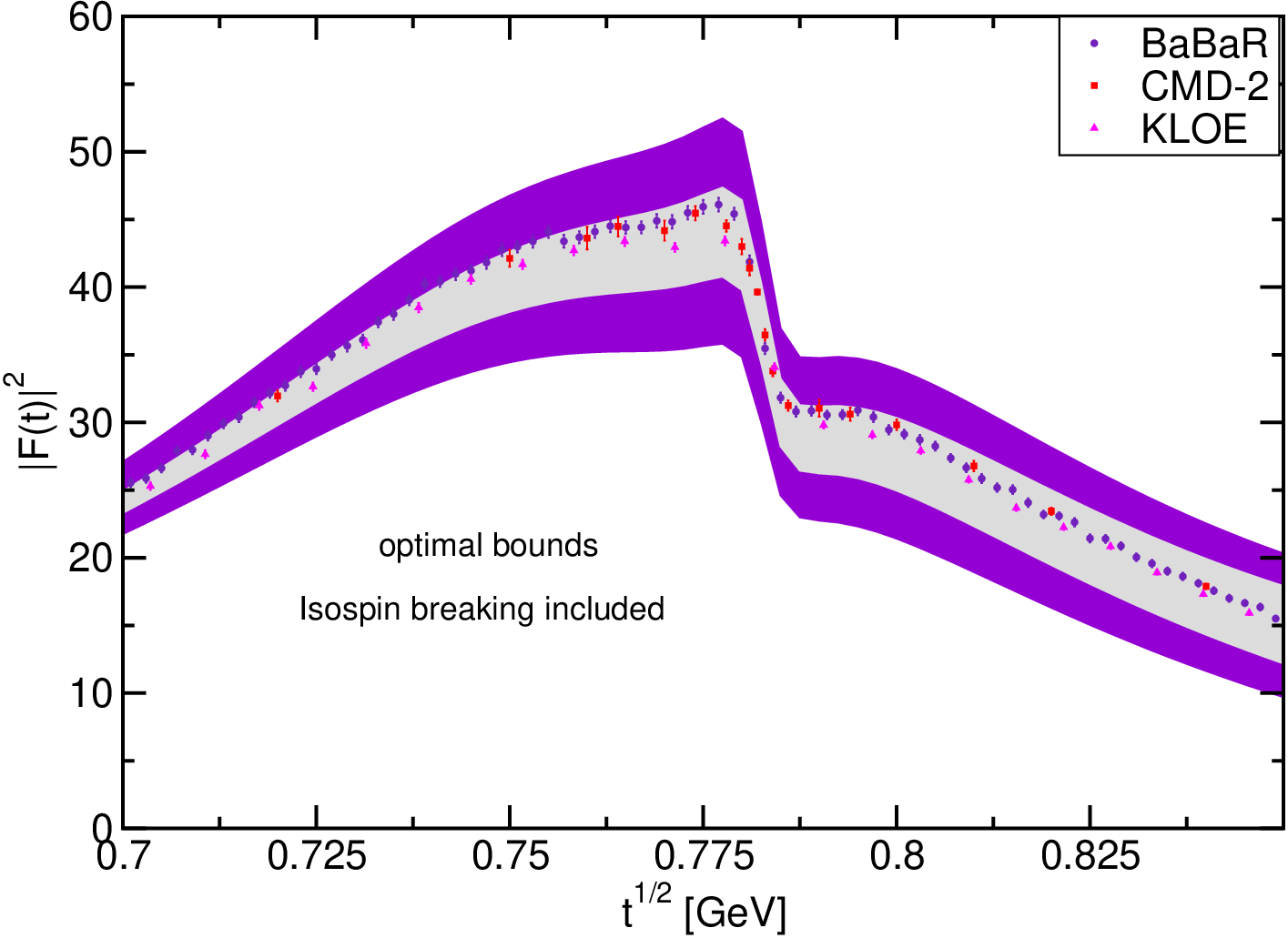}
\caption{Optimal bounds with isospin breaking correction  around the resonance region. The bands are as in Fig. \ref{fig:fig17}.}
\label{fig:fig22}
\end{figure}

Based on the above discussion, one can define "optimal" bounds by combining the results provided by different weights. For instance, we can take the upper bound given by   the weight $\rho(t) = 1/t$, and the lower bound 
given by  $\rho(t) = t^{1/2}/(t+3)$. These slightly improved bounds are presented in Figs. \ref{fig:fig17} - \ref{fig:fig22},
both in the isospin limit and  with isospin breaking corrections. We have performed the exercise  mainly for purposes of illustration, 
as the improvement is actually limited. Therefore, the  comparison with the experimental data 
\cite{BABAR,KLOE1,KLOE2,CMD2:1,CMD2:2,Fujikawa:2008ma} leads to conclusions similar to those already formulated for 
the specific weights: above the $\rho$ peak the data lie comfortably within the central band,  below the $\rho$ peak they  are near the upper edge of the central band,  while in the low-energy region the bounds 
are very strong, superseding  the  experimental data. 

\section{Discussion and Conclusions}\label{sec:conclusions}
Analyticity and unitarity have been used  in a large number of analyses of the pion electromagnetic form factor 
(for a list of references see \cite{Ananthanarayan:2012tn}).  A new formalism based on analyticity and unitarity, 
which  exploits  in an optimal way the precise information available on both the spacelike and the timelike axes, 
was  applied in \cite{Ananthanarayan:2011xt} and \cite{Ananthanarayan:2012tn} for constraining the shape of the 
form factor at $t=0$ and on the spacelike axis, respectively. In the present work we applied the same formalism 
for deriving upper and lower bounds on the 
modulus of the pion form factor on the unitarity cut below the $\omega\pi$ threshold. 

The phase and the modulus are related by analyticity, but a complete reconstruction of the modulus 
requires the exact knowledge of the phase along the  entire unitarity cut and the position of the possible zeros in 
the complex plane. As this information is not fully available, the correlation between the phase and the modulus  
is commonly investigated   by means of specific parametrizations, or by making {\it ad-hoc} assumptions in dispersive 
representations of the Omn\`es type.

Our aim was to check the consistency between the phase and the modulus of the pion  form factor in a model 
independent framework. We have used  as input  the phase in the elastic region, known via the Fermi-Watson theorem from 
the phase shift of the $P$-wave  of $\pi\pi$ scattering.  The lack of information on the phase above the inelastic 
threshold and
the zeros in the complex plane  has been compensated by  some experimental information and  conservative assumptions on the modulus above this threshold.   We have used the BaBar data  on the modulus \cite{BABAR} in the range between the $\omega\pi$  threshold and 3 GeV,
and very conservative assumptions  above 3 GeV.  Using modulus information above  $\tin$ and the phase below $\tin$, 
it is  possible to calculate upper and lower bounds on the modulus in the elastic region.

 In order to reduce the influence of the intermediate energies, where the modulus is not known, 
we implemented the information on $|F(t)|$ above  $\tin$  through the $L^2$ integral condition  
(\ref{eq:L2}), instead of imposing this knowledge pointwise, at each $t$.  Then the bounds are 
inevitably weaker, but for suitable choices of the weight $\rho(t)$ they are not sensitive to 
the assumptions on the modulus at intermediate and high energies.  We also included information 
on the normalization and charge radius at $t=0$ and the values at some spacelike momenta, available 
from recent experiments \cite{Horn,Huber}.

 We mention that, for a given input, the bounds are optimal and imply no loss of information as they are
obtained by solving exactly a functional interpolation problem. As seen from the formalism presented in 
Sec. \ref{sec:method}, the results  do not rely on specific parametrizations. Furthermore,  
the knowledge of the phase of the form factor in the inelastic region  is not required. Indeed, 
although some intermediate quantities entering the solution of the problem involve the phase chosen 
as an arbitrary (smooth) function above  $\tin$, it was proven  and checked numerically  
\cite{Abbas:2010EPJA} that the dependence is exactly compensated among the relevant factors.

We have compared the bounds obtained from our analysis with the precise 
measurements of the modulus in the elastic region 
from recent experiments on $e^+e^-$ annihilation 
\cite{BABAR,KLOE1,KLOE2,CMD2:1,CMD2:2} and 
hadronic $\tau$ decays \cite{Fujikawa:2008ma}. 
Of course, we could use as 
input also one of the measured values on the modulus below the inelastic threshold. This 
would lead to a considerable strengthening
of the bounds. However, as the four sets of data  are not 
perfectly consistent among them, the choice 
of the input from one set would introduce a certain bias, which we prefer to avoid in this 
first analysis. Since the problem is of great interest, it will be studied in the near future.

The bounds obtained from our analysis in the isospin symmetry limit  can be compared directly with the recent experimental 
data on the modulus available from hadronic $\tau$ decays \cite{Fujikawa:2008ma}. For the comparison with the data 
obtained from $e^+e^-$ annihilation \cite{BABAR,KLOE1,KLOE2,CMD2:1,CMD2:2}, isospin breaking was included in the formalism through the phase modification (\ref{eq:iso}).  In both cases, we have obtained a very good consistency 
above the $\rho$ peak, where the data lie well within the inner bands derived with the central values of our input. 

On the other 
hand, below the $\rho$ peak the data are situated at the upper edge of the central region. As this may indicate that our input is not optimal, we briefly analyzed in Sec. \ref{sec:results} the possible changes in the input that can improve the consistency.
The results show that it is not possible to achieve a very good consistency at all energies by tuning the value of the charge radius: with a slightly larger input value, $\la r_\pi^2 \ra =0.435 \,\mbox{fm}^2$ instead of the central value quoted in (\ref{eq:r2}), we obtain good consistency below and around the $\rho$ peak, but the agreement is worsened at higher energies.  
Furthermore, higher values of the charge radius are even less acceptable, 
as they lead to central bands in disagreement with the experimental data.

\bigskip

\vspace{2cm}

\bigskip

\bigskip

\bigskip

\vspace{-1cm}

Concerning the input phase, preliminary studies show that the bounds are quite sensitive to the detailed shape 
of the phase in the elastic region. We recall that the input phase was obtained via the Fermi-Watson theorem 
from the  $P$-wave phase shift of $\pi\pi$ scattering.  This phase shift was determined by solving Roy equations 
for $\pi\pi$ amplitude, using however as input the phase at a matching point from  phenomenological representations 
of the form factor. The phase used in the present analysis is taken from  
\cite{ACGL,CGL, KPY,GarciaMartin:2011cn,Caprini:2011ky}, which used previous measurements of the modulus 
(in particular, Refs. \cite{ACGL,CGL,Caprini:2011ky} took the input value of the $P$-wave phase shift 
at the matching point 0.8 GeV from a parametrization of CLEO data \cite{CLEO}).  Of course, one can 
adjust the input value until the phase obtained by solving Roy equations leads to upper and lower bounds 
consistent with the  more precise recent data on the modulus. This more elaborate analysis is also 
a project for future work. Taken together with other recent investigations \cite{Jegerlehner:2011ti,Benayoun:2011mm}, 
such a study will contribute to a better understanding of the form factor in the elastic region, 
crucial for the precise evaluation of the hadronic contribution to the anomalous magnetic
moment of the muon \cite{Jegerlehner:2009ry, Davier}.

The  considerations above,  based only on  the inner band,  
illustrate mainly the sensitivity of the results to the variation of the input. For checking the consistency we must of course take into account the enlarged bands, obtained by allowing the input to vary within errors. From the results presented in Section \ref{sec:results}  we conclude that there are no inconsistencies  between the bounds and the recent experimental data on the modulus above 0.5 GeV.

At low energies the bounds derived in the present work are very stringent, 
leading to allowed ranges for the modulus smaller  than the quoted errors of the experimental measurements. 
This is one of the most important result of our work, showing that the knowledge of the form factor at low energies, 
specifically below 0.5 GeV, can be improved  by exploiting in an optimal way  the phase in the elastic region 
and conservative information on the modulus above the inelastic threshold.  Also,  below 0.4 GeV  the results reveal some inconsistencies between the bounds and the experimental 
data on the modulus, especially from BaBar, despite the large errors on these data.  

We recall that the bounds are obtained with a definite range for the charge radius, given in (\ref{eq:r2}), and are sensitive to this input. As the adopted range (\ref{eq:r2}) leads to consistency for data on modulus at higher energies, but to some inconsistencies at low energies, the results indirectly signal some inconsistencies between data themselves.
As mentioned in the Introduction, the formalism  presented in this paper can  be exploited also
the other way round, for finding bounds on the 
pion charge  radius from the precise 
measurements on the modulus below the $\omega\pi$ threshold.
This analysis  will be performed in a future work. 

We finally mention that  further improvements are expected if  very precise determinations
of the form factor from ChPT, lattice simulations or experiment are made available. 
Indeed, as we have shown, precisely known input values  of the form factor inside the analyticity domain lead to 
a considerable strengthening  of the bounds. Therefore, the present formalism is a useful framework, open to further 
developments,  for analyticity tests and precise predictions on the pion form factor at low and intermediate energies.

\vskip0.2cm
\noindent{\bf Acknowledgement:}  We are grateful to H. Leutwyler for very useful comments and suggestions on the manuscript. 
This work was supported by Romanian CNCS in the Program Idei-PCE, Contract No. 121/2011.


\newpage

\begin{thebibliography}{100}


\bibitem{BABAR}
  B.~Aubert {\it et al.}  [BABAR Collaboration],
  Phys.\ Rev.\ Lett.\  {\bf 103} (2009) 231801
  [arXiv:0908.3589 [hep-ex]].

\bibitem{KLOE1}
  F.~Ambrosino {\it et al.}  [KLOE Collaboration],
  Phys.\ Lett.\ B {\bf 670} (2009) 285
  [arXiv:0809.3950 [hep-ex]].

\bibitem{KLOE2}
  F.~Ambrosino {\it et al.}  [KLOE Collaboration],
  Phys.\ Lett.\ B {\bf 700} (2011) 102
  [arXiv:1006.5313 [hep-ex]].

\bibitem{CMD2:1}
  R.~R.~Akhmetshin, V.~M.~Aulchenko, V.~S.~.Banzarov, L.~M.~Barkov, N.~S.~Bashtovoy, A.~E.~Bondar, D.~V.~Bondarev and A.~V.~Bragin {\it et al.},
  JETP Lett.\  {\bf 84} (2006) 413
   [Pisma Zh.\ Eksp.\ Teor.\ Fiz.\  {\bf 84} (2006) 491]
  [hep-ex/0610016].

\bibitem{CMD2:2}
  R.~R.~Akhmetshin {\it et al.}  [CMD-2 Collaboration],
  Phys.\ Lett.\ B {\bf 648} (2007) 28
  [hep-ex/0610021].

\bibitem{Fujikawa:2008ma}
  M.~Fujikawa {\it et al.}  [Belle Collaboration],
  Phys.\ Rev.\ D {\bf 78} (2008) 072006
  [arXiv:0805.3773 [hep-ex]].

\bibitem{ACGL}
  B.~Ananthanarayan, G.~Colangelo, J.~Gasser and H.~Leutwyler,
  Phys.\ Rept.\  {\bf 353} (2001) 207
  [hep-ph/0005297].

\bibitem{CGL}
  G.~Colangelo, J.~Gasser and H.~Leutwyler,
  Nucl.\ Phys.\ B {\bf 603} (2001) 125
  [hep-ph/0103088].

\bibitem{KPY}
  R.~Kaminski, J.~R.~Pelaez and F.~J.~Yndurain,
  Phys.\ Rev.\ D {\bf 77} (2008) 054015
  [arXiv:0710.1150 [hep-ph]].

\bibitem{GarciaMartin:2011cn}
  R.~Garcia-Martin, R.~Kaminski, J.~R.~Pelaez, J.~Ruiz de Elvira and F.~J.~Yndurain,
  Phys.\ Rev.\ D {\bf 83} (2011) 074004
  [arXiv:1102.2183 [hep-ph]].

\bibitem{Caprini:2011ky}
  I.~Caprini, G.~Colangelo and H.~Leutwyler,
  Eur.\ Phys.\ J.\ C {\bf 72} (2012) 1860
  [arXiv:1111.7160 [hep-ph]].

\bibitem{Colangelo:2004}
  G.~Colangelo,
  Nucl.\ Phys.\ Proc.\ Suppl.\  {\bf 131} (2004) 185
  [hep-ph/0312017].

\bibitem{Masjuan}
  P.~Masjuan, S.~Peris and J.~J.~Sanz-Cillero,
  Phys.\ Rev.\ D {\bf 78} (2008) 074028
  [arXiv:0807.4893 [hep-ph]].

\bibitem{Horn}
  T.~Horn {\it et al.}  [Jefferson Lab F(pi)-2 Collaboration],
  Phys.\ Rev.\ Lett.\  {\bf 97} (2006) 192001
  [nucl-ex/0607005].

\bibitem{Huber}
  G.~M.~Huber {\it et al.}  [Jefferson Lab Collaboration],
  Phys.\ Rev.\ C {\bf 78} (2008) 045203
  [arXiv:0809.3052 [nucl-ex]].

\bibitem{IC}
  I.~Caprini,
  Eur.\ Phys.\ J.\ C {\bf 13} (2000) 471
  [hep-ph/9907227].

\bibitem{Abbas:2010EPJA}
  G.~Abbas, B.~Ananthanarayan, I.~Caprini, I.~Sentitemsu Imsong and S.~Ramanan,
  Eur.\ Phys.\ J.\ A {\bf 45} (2010) 389
  [arXiv:1004.4257 [hep-ph]].

\bibitem{AC}
  B.~Ananthanarayan and I.~Caprini,
  J.\ Phys.\ Conf.\ Ser.\  {\bf 374} (2012) 012011
  [arXiv:1202.5391 [hep-ph]].

\bibitem{Ananthanarayan:2011xt}
  B.~Ananthanarayan, I.~Caprini and I.~S.~Imsong,
  Phys.\ Rev.\ D {\bf 83} (2011) 096002
  [arXiv:1102.3299 [hep-ph]].

\bibitem{Ananthanarayan:2012tn}
  B.~Ananthanarayan, I.~Caprini and I.~S.~Imsong,
  Phys.\ Rev.\ D {\bf 85} (2012) 096006
  [arXiv:1203.5398 [hep-ph]].


\bibitem{Cornille:1975} H. Cornille and A. Martin,  Nucl. Phys. B {\bf 93},  61 (1975).

\bibitem{Leutwyler:2002hm}
  H.~Leutwyler,
  hep-ph/0212324.

\bibitem{Jegerlehner:2011ti}
  F.~Jegerlehner and R.~Szafron,
  Eur.\ Phys.\ J.\ C {\bf 71} (2011) 1632
  [arXiv:1101.2872 [hep-ph]].

\bibitem{Benayoun:2011mm}
  M.~Benayoun, P.~David, L.~DelBuono and F.~Jegerlehner,
  Eur.\ Phys.\ J.\ C {\bf 72} (2012) 1848
  [arXiv:1106.1315 [hep-ph]].

\bibitem{Jegerlehner:2009ry}
  F.~Jegerlehner and A.~Nyffeler,
  Phys.\ Rept.\  {\bf 477} (2009) 1
  [arXiv:0902.3360 [hep-ph]].

\bibitem{Davier}	M. Davier, A. Hoecker, B. Malaescu and Z. Zhang 
Eur. Phys. J. C {\bf 71} (2011) 1515, Erratum-ibid. C {\bf 72} (2012) 1874
[arXiv:1010.4180 [hep-ph]].
 
\bibitem{Hanhart:2012wi}
  C.~Hanhart,
Phys. Lett. B{\bf 715}, 170 (2012)
  arXiv:1203.6839 [hep-ph].

\bibitem{CLEO}
S. Anderson {\it et al.}  [CLEO Collaboration], Phys. Rev. D{\bf 61} (2000) 112002 [hep-ex/9910046].
\end{thebibliography}
\end{document}